\documentclass[aps,pra,twocolumn,nofootinbib,preprintnumbers]{revtex4-1}

\usepackage{amsmath,amssymb,amsfonts,graphicx}
\usepackage{bm}
\usepackage{color}
\usepackage{subfigure}
\usepackage[hidelinks]{hyperref}
\usepackage{latexsym}
\usepackage{multirow}
\usepackage{mathrsfs}
\usepackage{feynmf}

\renewcommand{\vec}[1]{{\mathbf{#1}}}
\newcommand{\beq}{\begin{eqnarray}}
\newcommand{\eeq}{\end{eqnarray}}

\newcommand{\tabii}{\hspace{.2\textwidth}}

\newcommand{\tabiv}{\hspace{.4\textwidth}}

\begin{document}

\title{Realizing infrared power-law liquids in the cuprates from unparticle interactions}
 \author{Kridsanaphong Limtragool, Chandan Setty, Zhidong Leong, and Philip W. Phillips}
\affiliation{Department of Physics and Institute for Condensed Matter Theory,
University of Illinois
1110 W. Green Street, Urbana, IL 61801, U.S.A.}
\date{\today}

\begin{abstract}
Recent photoemission experiments \cite{dessau} reveal that the excitations along the nodal region in the strange metal of the cuprates, rather than corresponding to poles in the single-particle Green function, exhibit power-law scaling as a function of frequency and temperature.  Because such power-law scaling is indicative of a scale-invariant sector, as a first step, we perturbatively evaluate the electron self-energy due to interactions with scale-invariant unparticles.  We focus on a $G_0W$ type diagram with an interaction $W$ mediated by a bosonic scalar unparticle. We find that, in the high-temperature limit, the imaginary part of the self-energy $\mathrm{Im}\Sigma$ is linear in temperature. In the low-temperature limit, $\mathrm{Im}\Sigma$ exhibits the same power law in both temperature and frequency, with an exponent that depends on the scaling dimension of an unparticle operator. Such behavior is qualitatively consistent with the experimental observations.  We then expand the unparticle propagator into coherent and incoherent contributions, and study how the incoherent part violates the density of states (DOS) and density-density correlation function sum rules (f-sum rule).  Such violations can, in principle,  be observed experimentally.  Our work indicates that the physical mechanism for the origin of the power-law scaling is the incoherent background, which is generated from the Mott-scale physics.
\end{abstract}
\maketitle

\section{Introduction}

Scale invariance is the cornerstone of any infrared theory or physical system in which the physics is controlled by a critical fixed point.  It is surprising then that recent photoemission experiments \cite{dessau} on the strange metal phase of the cuprates indicate that scale invariance is applicable not just at a single (quantum critical) point, but over an entire phase. The experiments reveal that the excitations along the nodal direction are not at all quasiparticle-like, but rather exhibit a power-law scaling in which frequency and temperature are interchangeable. Specifically, they find \cite{dessau} that the scattering rates inferred from the momentum distribution curves (MDCs) along the nodal direction exhibit a power-law scaling of the form,
\beq
\Sigma_{\rm PLL}(\omega)=\Gamma_0+\lambda\frac{((\hbar\omega)^2+(\gamma k_B T)^2)^a}{(\hbar\omega_N)^{2a-1}},
\eeq
where $\omega_N$ is a normalization energy scale, and  $\gamma$ and $\lambda$ are constants.  This scaling form of the self-energy, motivated by that of marginal Fermi liquid theory 
\cite{mfl} and Anderson's \cite{anderson} ``Hidden Fermi Liquid'', is found to hold over a wide range of dopings --- from underdoped where $a<1/2$, to optimal doping where $a=1/2$,  and to overdoped with $a>1/2$. While scaling has been observed previously, it was typically associated with just a single doping level, namely at the optimal concentration \cite{Marel2003,Valla1999,Abrahams2000}.  

It is the modeling and possible origin of such power-law scaling over an entire phase that we address here.   Scale invariance over a wide range does have some precedence, at least theoretically. For example, the spectral functions calculated within the AdS/CFT formalism exhibit a  range of power law scaling as the scaling dimension of the boundary fermionic operator is tuned continuously \cite{faulkner,zaanen}.  Building on this set of ideas, one of us \cite{Phillips2013} proposed that the strange metal is described by a scale-invariant sector over the entire range of doping.  The key computational tool used here is the unparticle propagators proposed by Georgi \cite{Georgi2007}.   The unparticle propagators were modeled within the continuous mass formalism.  If $\phi( p, m^2)$ is a scalar propagator with four momentum $\textit p$ and mass squared $m^2$, the propagator for the unparticle ($G_u(p)$) can be obtained as \cite{Krasnikov2007,Deshpande2008} $G_u(p) = \int_0^{\infty} dm^2 \phi(p,m^2) f^2(m^2)$ with $ \phi(p,m^2) = \frac{1}{p^2 - m^2 + i \eta} $. The function $f(m^2)$ is a weighing function such that the number of fields between $m^2$ and $m^2+dm^2$ is $f(m^2)dm^2$. If we choose  $f(m^2) = (m^2)^{(d_u - \frac{d+1}{2})/2} $ where $d +1$ is the spacetime dimension and $d_u$ is the scaling dimension of the unparticle operator, we obtain the unparticle propagator
\begin{equation}\label{UnparticlePropagator}
G_u(p) \sim \frac{1}{(-p^2 - i \eta)^{\frac{d+1}{2} - d_u}}.
\end{equation}
Alternatively, to construct unparticles, one can start from the action of a massive scalar field \cite{Phillips2013} and integrate over the mass $m^2$ (again assuming a mass distribution) by treating it as an additional coordinate so that scale invariance is restored. The new action can then be rewritten into a theory in the anti-de Sitter space-time (AdS). The resulting two point function obtained from the AdS/CFT correspondence can be identified as the unparticle propagator. It is clear from the form of the unparticle propagator that the exponent $\frac{d+1}{2} - d_u$ is, in general, not an integer. This yields branch points at $\pm \textbf k^2$ (instead of poles), indicating that unparticles represent incoherent electronic states of matter that lack a ``particle-like'' character, and are associated with measurable quantities that encompass the physics from both low and high energy scales.  

A question that remains open from this work is how the unparticle sector interact with the particle sector to renormalize the quasiparticle weights.  Physically, the unparticle sector should be thought of as the incoherent part of the spectrum.  The question we address here is:  what is the fate of particles in the presence of an incoherent sector? This question is of utmost relevance at present since the experiments indicate that it is the electron scattering rate that exhibits a power-law scaling.  Ultimately, a secondary question that rises is what role do unparticle-unparticle interactions play.   We postpone the latter question to a subsequent paper and focus here on the former to see how close we can come to a description of the power-law liquid with just electron-unparticle interactions.  Consequently, to address the experiments, we consider the interaction between electrons and the unparticle sector directly.  The quantity we focus on is a quasiparticle's lifetime $\tau$, which is proportional to the imaginary part of electron's self-energy. In particular, we want to know how $\tau$ depends on the temperature, $T$, and frequency, $\omega$, and in what situation $\tau$ exhibits a power law as  a result of interactions with the scale-invariant sector.   We evaluate electron self-energies perturbatively using a $G_0W$ type diagram in which the interaction $W$ is mediated by the bosonic scalar unparticle sector.  We find that at high temperatures, the quasiparticle's lifetime is linear in $T$ as a result of bosonic excitations of unparticles. In the low-temperature limit, the electron's energy dispersion becomes linear in momentum and thus the scaling analysis can be applied. The quasiparticle's lifetime in this case is a power law of the form $\tau \sim T^{d-2+2\alpha}$ and $\tau \sim |\omega|^{d-2+2\alpha}$, where $1-\alpha = \frac{d+1}{2} -d_u$.  To satisfy the unitarity bound\cite{Minwalla1997} $d_u> (d-1)/2$, the only constraint on $\alpha$ is $\alpha>0$.  However, in perturbation theory, further constraints arising from the Matsubara summations and the convergence of the integrals place $\alpha$ in the interval $(3-d)/2<\alpha<1$.   Hence,  while the current perturbative particle-unparticle treatment can describe non-trivial power-law behaviour of the self-energy with respect to temperature and frequency, it cannot access the regime $\alpha< (3-d)/2$ where the current theory gives infrared divergences.  Whether the divergences which arise for $\alpha<(3-d)/2$ vanish when unparticle-unparticle interactions are included will be explored in a further publication.  We then turn our attention to a problem regarding a violation of a sum rule when unparticles are present in a system. In this discussion, instead of using the bosonic unparticles, we use a fermionic propagator with a fractional power, since the results obtained can be readily compared with known standard sum rules. We expand the propagator into coherent and incoherent contributions, and study how the incoherent part violates the density of states (DOS) and density-density correlation function sum rules. 

\section{Electron-Unparticle} \label{sec:elec-unpar}

\begin{figure}[h]
	\centering 
	\includegraphics[scale=0.5]{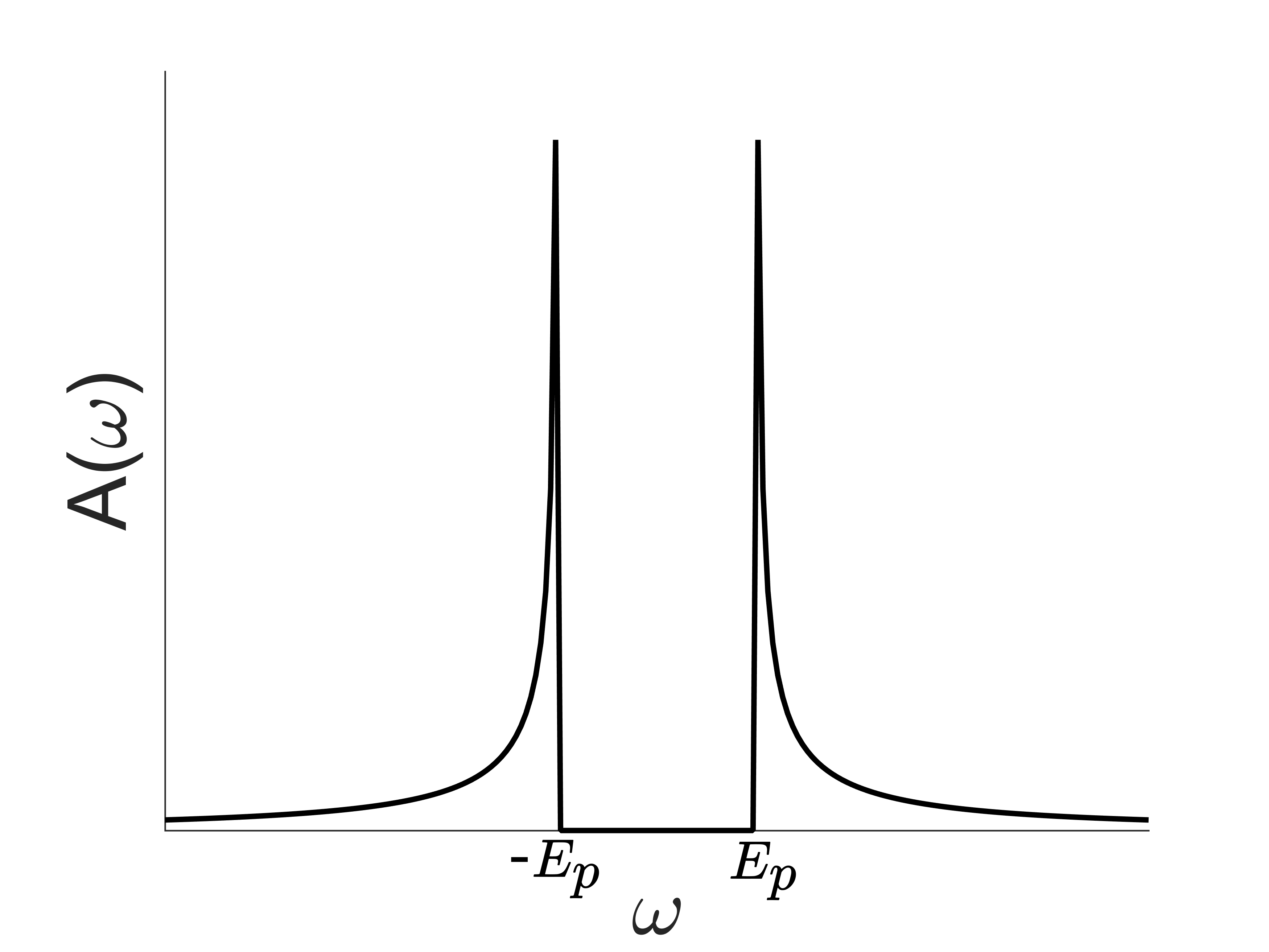}
    \caption{The spectral function of unparticles when $0<\alpha<1$.}  \label{fig:spectral}
\end{figure}

We investigate a system consisting of electrons that interact with a scale invariant sector. We model such a scale invariant sector by a bosonic scalar unparticle with a momentum cutoff $\Lambda$. The cutoff signifies that the unparticle is an effective infrared description of some high energy model. The interaction between an electron and an unparticle is chosen to be a constant Yukawa coupling, $u$. The action of the model we consider in Matsubara-Fourier space is given by
\beq
S &=& T\sum\limits_{m}\int\limits^{\Lambda}\frac{d^{d}p}{(2\pi)^{d}} \phi_{m}(\vec p)  G^{-1}_{u,m}(\vec p)\phi_{m}(\vec p) \nonumber \\
&& - T\sum\limits_{n}\int\limits\frac{d^{d}q}{(2\pi)^{d}}  \psi^\dagger_n(\vec q)G^{-1}_{e,n}(\vec q) \psi_n(\vec q)  \nonumber \\
&& +uT^2\sum\limits_{m,n}\int\limits^{\Lambda}\frac{d^dp}{(2\pi)^d}\int\limits\frac{d^dq}{(2\pi)^d}\psi^\dagger_{m+n} (\vec p+ \vec q)\phi_m(\vec p)\psi_n(\vec q)  \nonumber \\
\eeq
where $\phi$ is a bosonic unparticle and $\psi$ is a non-relativistic electron field. Here, $G_u$ is the unparticle propagator, $G_e$ is the non-interacting electron propagator, $T$ denotes the temperature, and the subscripts of the fields and the propagators denote the dependence on Matsubara frequency. The bosonic unparticle propagator is given by
\beq \label{eq:Gu}
G_{u,m}(\vec p) = \frac{1}{(\omega_m^2+E^2_{\vec p})^{1-\alpha}},
\eeq
where $\alpha$ is related to $d$ and $d_u$ by $1-\alpha = \frac{d+1}{2} - d_u$, and $E_{\vec p}$ is a quantity with units of energy. Since the spectral function calculated from $G_u$ is gapped between $-E_{\vec p}$ and $E_{\vec p}$ (Fig. \ref{fig:spectral}), $E_{\vec p}$ can be interpreted as the minimum energy required to excite an unparticle of momentum $p$. 

We choose $E_{\vec p}$ to have a form $E_{\vec p} = |\vec p|v = pv$ in order for $G_u$ to be scale covariant. Here, $v$ is a dimensionless constant. If $\alpha = 0$, $G_u$ turns into a propagator of a free scalar field. The electron propagator $G_e$ is given by
\beq
G_{e,n}(\vec q) = \frac{1}{i\omega_n - \varepsilon_{\vec q}},
\eeq
where $\varepsilon_{\vec p} = \frac{{p}^2}{2m} - \mu$ and $\mu=\varepsilon_f$ is the chemical potential which we assume to be temperature independent.

We want to point out that the system of electrons and unparticles we consider here resembles the standard electron-phonon system modeled by the Fr{\"o}hlich Hamiltonian \cite{Mahan2000}, but with key differences. In contrast to the phonon being a quanta of lattice vibration, the unparticle in our model is an effective scale invariant object of some high energy model, such as the Mott physics in the Hubbard model.   It is the lack of the quanta concept being relevant to unpartices that is the origin of the branch-cut behavior of the unparticle propagator. Nonetheless, as with phonons, unparticle stuff {\it exists} only up to $\Lambda$ in the same way that phonon has a momentum cutoff $\sim 1/a$ with $a$ being a lattice spacing.  

\begin{figure}[h]
	\centering
	\includegraphics[scale=0.4]{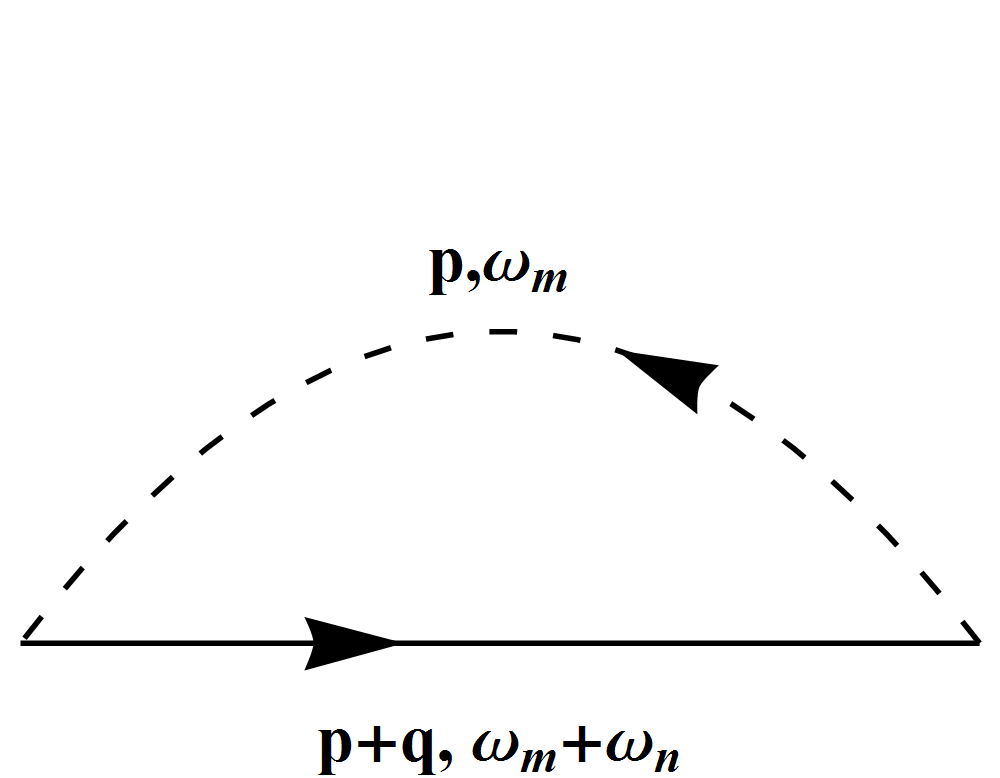}
	\caption{The Feynman diagram for the electron's self-energy. The solid line represents an electron propagator. The dashed line represents an unparticle propagator.} \label{fig:feynman}
\end{figure}

We are interested in the quasiparticle's lifetime $\tau$ which is given by $\tau = -\frac{1}{2}\mathrm{Im}\Sigma$, where $\Sigma$ is the electron's self-energy. The expression for the electron self-energy at the lowest order (Fig. \ref{fig:feynman}) as a function of fermionic Matsubara frequency $\omega_n$ and momentum $q$ can be written as
\beq \label{eq:sigma}
\Sigma_n(\vec q) &=& u^2\int \frac{d^dp}{(2\pi)^d}T\sum\limits_{m} G_{u,m}(\vec p)G_{e,n+m}(\vec q + \vec p).
\eeq
We perform the summation over the bosonic Matsubara frequency $\omega_m$ using the standard contour integral technique (see Appendix \ref{app:matsum}). We then perform an analytic continuation, $i\omega_n \rightarrow \omega + i\eta$, to obtain the retarded self-energy. The self-energy in the case of $0<\alpha<1$ is
\begin{widetext}
\beq \label{eq:sigma_total_1}
\Sigma(\omega,\vec q) &=& -u^2 \int \frac{d^dp}{(2\pi)^d} \frac{g_F(\varepsilon_{\vec p+ \vec q})}{(-\varepsilon_{\vec p+ \vec q}+E_{\vec p}+\omega+i\eta)^{1-\alpha}(\varepsilon_{\vec p+ \vec q}+E_{\vec p}-\omega-i\eta)^{1-\alpha}} \nonumber \\
&& + \frac{\sin(\pi\alpha)}{\pi}u^2 \int\frac{d^dp}{(2\pi)^d}\int\limits_{E_{\vec p}}^{\infty}dz \frac{g_B(z)}{(z+E_{\vec p})^{1-\alpha}(z-E_{\vec p})^{1-\alpha}}\bigg(\frac{1}{(z+\omega+i\eta-\varepsilon_{\vec p+ \vec q})} - \frac{1}{(z-\omega-i\eta+\varepsilon_{\vec p+ \vec q})}\bigg), 
\eeq
\end{widetext}
where $g_F(z) = \frac{1}{2}\tanh(\frac{\beta z}{2})$ is a fermionic factor and $g_B(z) = \frac{1}{2}\coth(\frac{\beta z}{2})$ is a bosonic factor obtained from converting the summation to the contour integral. Here, the phase angle needed when one evaluates the power $1-\alpha$ in the first term is in the range $-\pi\le\theta<\pi$. We denote the first term by $\Sigma_F$ and the second term by $\Sigma_B$ since their integrands contain the fermionic and bosonic factors. We are interested in the behavior of the imaginary part of this self-energy as a function of temperature and frequency. 

\subsection{Behavior of $\mathrm{Im}\Sigma(T)$} \label{sec:imsig_t}

We now turn to the evaluation of the relevant terms.  When $T\gg |\varepsilon_{\vec p + \vec q}|$, we have $g_F(\varepsilon_{\vec p + \vec q}) = \frac{1}{2}\tanh\frac{\beta\varepsilon_{\vec p+ \vec q}}{2} \approx \frac{\beta\epsilon_{\vec p+ \vec q}}{4}$.
Hence, the first term in Eq. \ref{eq:sigma_total_1}, $\Sigma_F$, goes like $O(\frac{1}{T})$. Taking the imaginary part of the second term in Eq. \ref{eq:sigma_total_1} and then integrating over $z$ using the delta functions, one obtains
\begin{widetext}
\beq \label{eq:sigma_b_int_z}
\mathrm{Im} \ \Sigma_{B}(\omega,\vec q) &=& -\sin(\pi\alpha)u^2\int\frac{d^dp}{(2\pi)^d} \Bigg(\Theta(-\omega+\varepsilon_{\vec p + \vec q}-E_{\vec p})\frac{g_B(-\omega+\varepsilon_{\vec p + \vec q})}{(-\omega+\varepsilon_{\vec p + \vec q}+E_{\vec p})^\alpha(-\omega+\varepsilon_{\vec p + \vec q}-E_{\vec p})^\alpha} \nonumber \\
&& \tabii + \Theta(\omega-\varepsilon_{\vec p + \vec q}-E_{\vec p})\frac{g_B(\omega-\varepsilon_{\vec p + \vec q})}{(\omega-\varepsilon_{\vec p + \vec q}+E_{\vec p})^\alpha(\omega-\varepsilon_{\vec p + \vec q}-E_{\vec p})^\alpha} \Bigg).
\eeq
\end{widetext}
In the high-temperature limit, $g_B(z)$ can be expanded as $g_B(z) = \frac{1}{2}\tanh\frac{\beta z}{2}\rightarrow \frac{1}{\beta z} + O(\beta z).$
So when $T\gg |\epsilon_{\vec p + \vec q}-\omega|$, $\mathrm{Im} \Sigma_{B}(\omega,\vec q) = -Cu^2T + O(\frac{1}{T})$ where $C$ is a temperature-independent constant. Consequently, in the high-temperature limit, one has
\beq
\mathrm{Im}\Sigma = \mathrm{Im}\Sigma_F + \mathrm{Im}\Sigma_B \propto -T.
\eeq
Since we do not explicitly use the form $E_{\vec p} = pv$ in the above argument, this result also holds for any form of $E_{\vec p}$, provided the integral in Eq. \ref{eq:sigma_b_int_z} converges.

\begin{figure}[h]
	\centering 
	\includegraphics[scale=0.5]{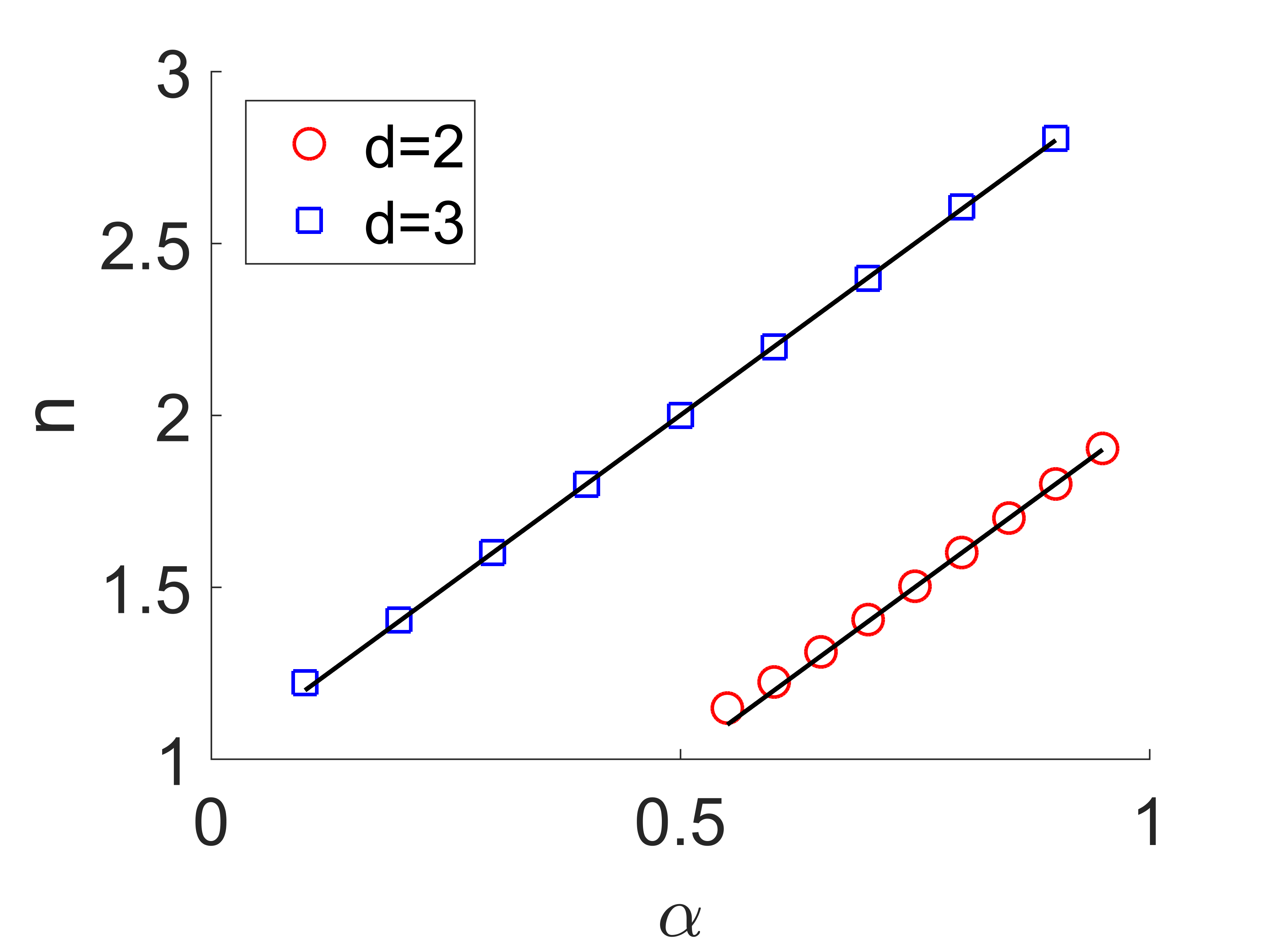}
    \caption{Plots of $\mathrm{Im}\Sigma$'s temperature power law exponent $n$ vs. $\alpha$ at low temperatures. Squares and circles correspond to the exponents obtained by fitting the low $T$ parts of $\mathrm{Im}\Sigma$ to power law in $d=3$ and $d=2$, respectively. Black lines are the plots of $n = d-2+2\alpha$.}  \label{fig:lowTpower}
\end{figure}

For the low-temperature case, we consider only the electrons on the Fermi surface ($\vec q = \vec q_f$) with $\omega = 0$. In this case, the expression $\varepsilon_{\vec p + \vec q}$ simplifies to $\varepsilon_{\vec p + \vec q} = \frac{{p}^2}{2m} + \frac{\vec p\cdot \vec q}{m}$. If the momentum cutoff $\Lambda$ is much smaller than $q_f$, it should be reasonable to omit the term $\frac{{p}^2}{2m}$ in $\varepsilon_{\vec p + \vec q}$. From Eq. \ref{eq:sigma_total_1}, we separate the temperature dependent parts  using $g_F(z) = \frac{1}{2} - n_F(z)$ and $g_B(z) = \frac{1}{2} + n_B(z)$ where $n_F$ and $n_B$ are the Fermi and Bose distributions, respectively. We drop the $\frac{1}{2}$ terms from $g_F$ and $g_B$, since we are only interested in the temperature dependence of $\Sigma$. Performing the change of variables $z\rightarrow pz'$ and $p\rightarrow pT$, one obtains
\begin{widetext}
\beq
\Sigma &=& u^2T^{d-2+2\alpha} \int\limits^{\Lambda/T} \frac{d^dp}{(2\pi)^d} p^{-2+2\alpha}\frac{1}{(e^{\frac{p\cdot \vec q_f}{m}}+1)(-\frac{\hat{p}\cdot \vec q_f}{m}+v+i\eta)^{1-\alpha}(\frac{\hat{p}\cdot \vec q_f}{m}+v-i\eta)^{1-\alpha}} \nonumber \\
&& + \frac{\sin(\pi\alpha)}{\pi}u^2T^{d-2+2\alpha} \int\limits^{\Lambda/T}\frac{d^dp}{(2\pi)^d}p^{-2+2\alpha}\int\limits_{v}^{\infty}dz' \frac{1}{(e^{pz'}-1)(z'+v)^{1-\alpha}(z'-v)^{1-\alpha}}\bigg(\frac{1}{(z'+i\eta-\frac{\hat{p}\cdot \vec q_f}{m})} - \frac{1}{(z'-i\eta+\frac{\hat{p}\cdot \vec q_f}{m})}\bigg) \nonumber \\
\eeq
\end{widetext}
where $\hat{p}$ denotes a unit vector in the direction of $\vec p$. Upon taking the limit $T \rightarrow 0$, the upper limit of the momentum integral can be taken to $\infty$ provided that there is no infrared divergence from the integrals over $p$. By counting the powers of $p$, one needs $d-3+2\alpha > 0$, i.e. $\alpha>0$ for $d = 3$ and $\alpha > 0.5$ for $d = 2$. Therefore, one has
\beq \label{eq:imsig_lowT}
\mathrm{Im}\Sigma \propto -T^{d-2+2\alpha}
\eeq
at low temperatures. If the coupling constant $u$ depends on momentum $\vec p$, the criterion for the absence of the infrared divergence and the scaling of $\mathrm{Im}\Sigma$ will be different. A more in-depth analysis for the $d = 3$ case in Appendix \ref{app:scaling_imsigma} shows that when $\frac{\Lambda}{2} < q_f - mv$ and $mv<q_f$ the term $\frac{p^2}{2m}$ in $\varepsilon_{\vec p + \vec q}$ contributes to $\mathrm{Im}\Sigma$ as $O(T^{d-1+2\alpha})$. This results justifies our omission of the $\frac{p^2}{2m}$ term in the scaling analysis above. The same argument used in Appendix \ref{app:scaling_imsigma} cannot be applied to the $d = 2$ case. Nevertheless, the numerical results below indicate that Eq. \ref{eq:imsig_lowT} still holds for the case $d = 2$ and $\alpha>0.5$ (see Fig. \ref{fig:lowTpower}).

\begin{figure}[h!]
	\centering 
		\subfigure[\ $d = 3$ and $\alpha = 0.8$ \label{fig:sigma3d}]{\includegraphics[scale=0.5]{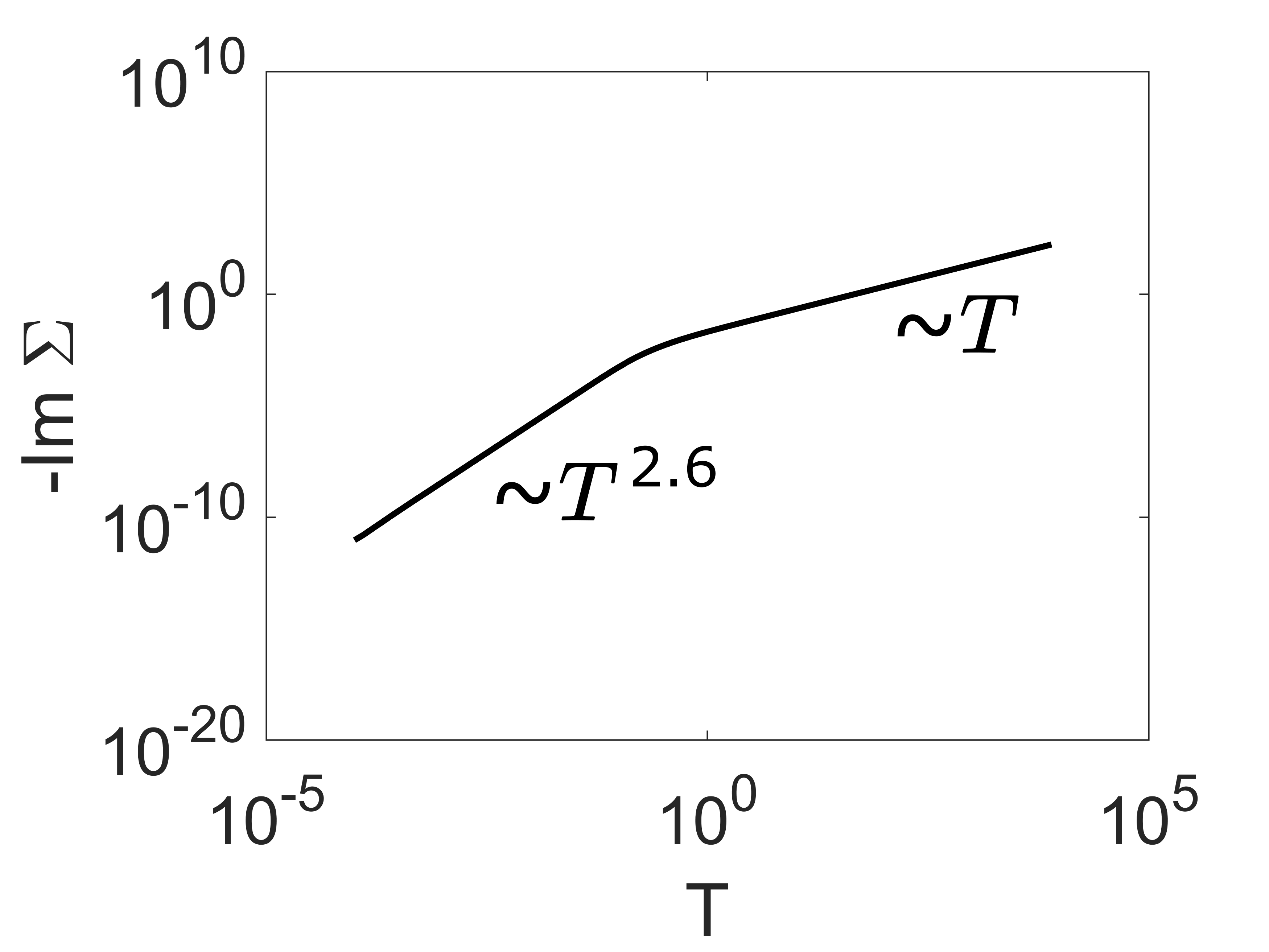}} 
		\subfigure[\ $d = 2$ and $\alpha = 0.8$  \label{fig:sigma2d}]{\includegraphics[scale=0.5]{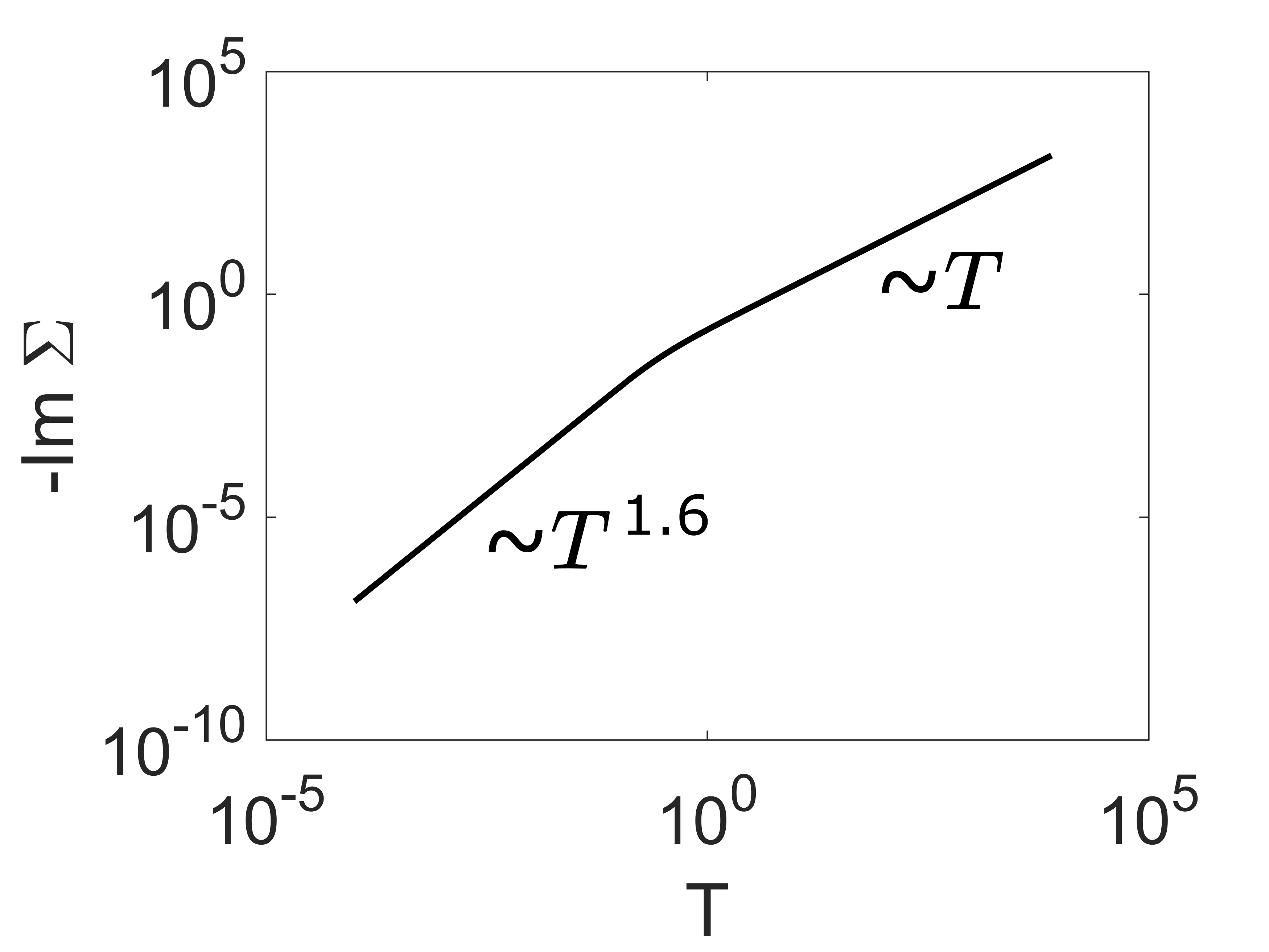}}
    \caption{Log-log plots of the imaginary part of the self-energy as a function of temperature.} 
\end{figure}

We numerically evaluate the imaginary part of the self-energy as a function of temperature using Eq. \ref{eq:sigma_combine_final_3d} for $d=3$ and Eq. \ref{eq:sigma_combine_final_2d} for $d=2$. Here we use $\Lambda = m$, $q_f = \sqrt{2}m$, and $v = 0.4$. With these parameters, the conditions $\frac{\Lambda}{2} < q_f - mv$ and $mv<q_f$ are satisfied. The results for the $\alpha = 0.8$ case are shown in Figs. \ref{fig:sigma3d} and \ref{fig:sigma2d}. We find that, for both the $d=3$ and $d=2$ cases, $\mathrm{Im}\Sigma$ depends linearly on temperature at large $T$. At low $T$, $\mathrm{Im}\Sigma$ exhibits a power law $\sim T^n$. The exponent $n$ follows Eq. \ref{eq:imsig_lowT} for $d = 3$ when $0<\alpha<1$, and for $d = 2$ when $0.5<\alpha<1$ as shown in Fig. \ref{fig:lowTpower}.

\subsection{Behavior of $\mathrm{Im}\Sigma(\omega)$}

\begin{figure}[h]
	\centering 
	\subfigure[\ Low temparture $T = 0.01m$ \label{fig:imsig_w_lowT}]{\includegraphics[scale=0.45]{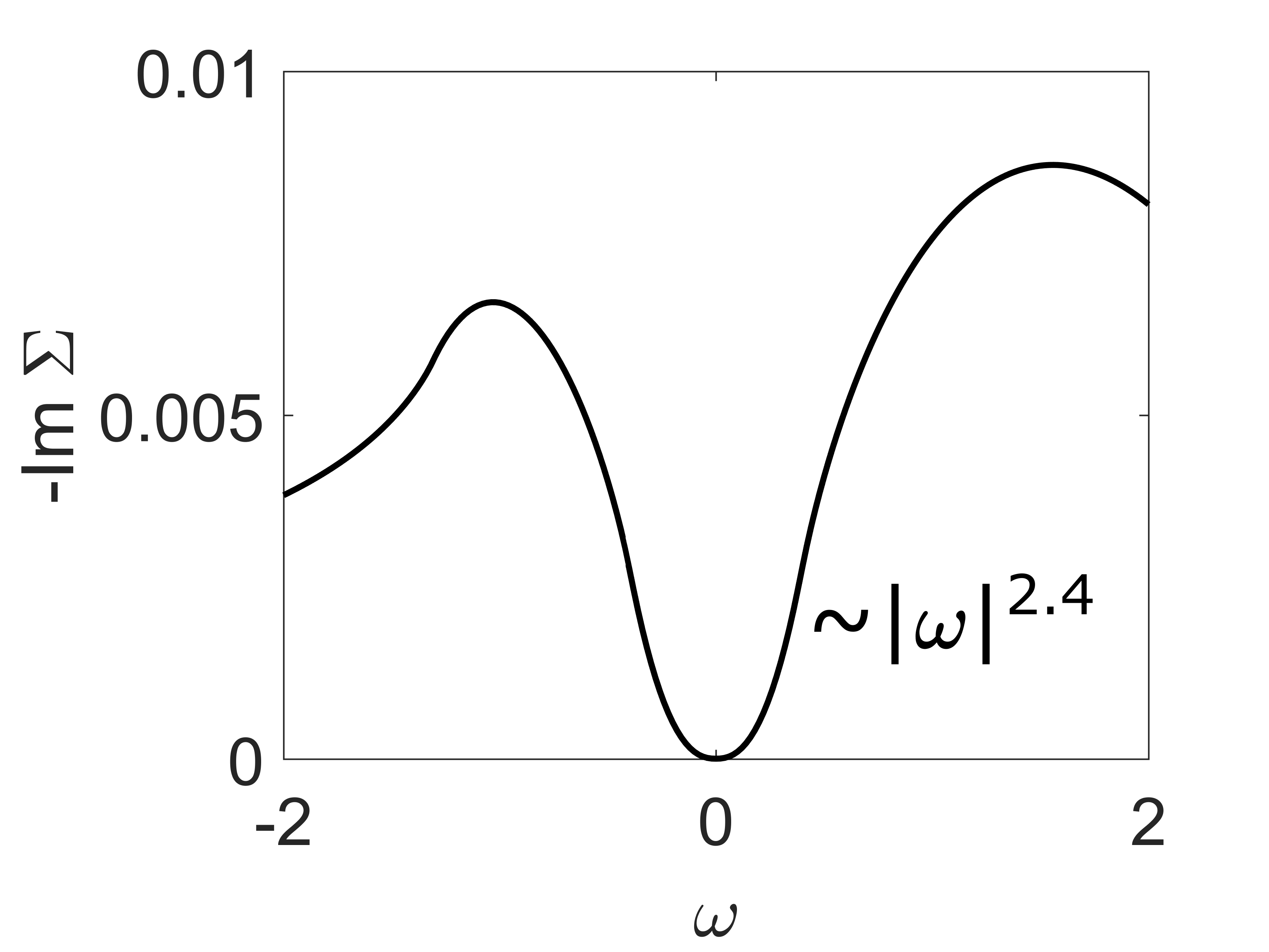}} 
	\subfigure[\ High temperature $T = m$  \label{fig:imsig_w_largeT}]{\includegraphics[scale=0.45]{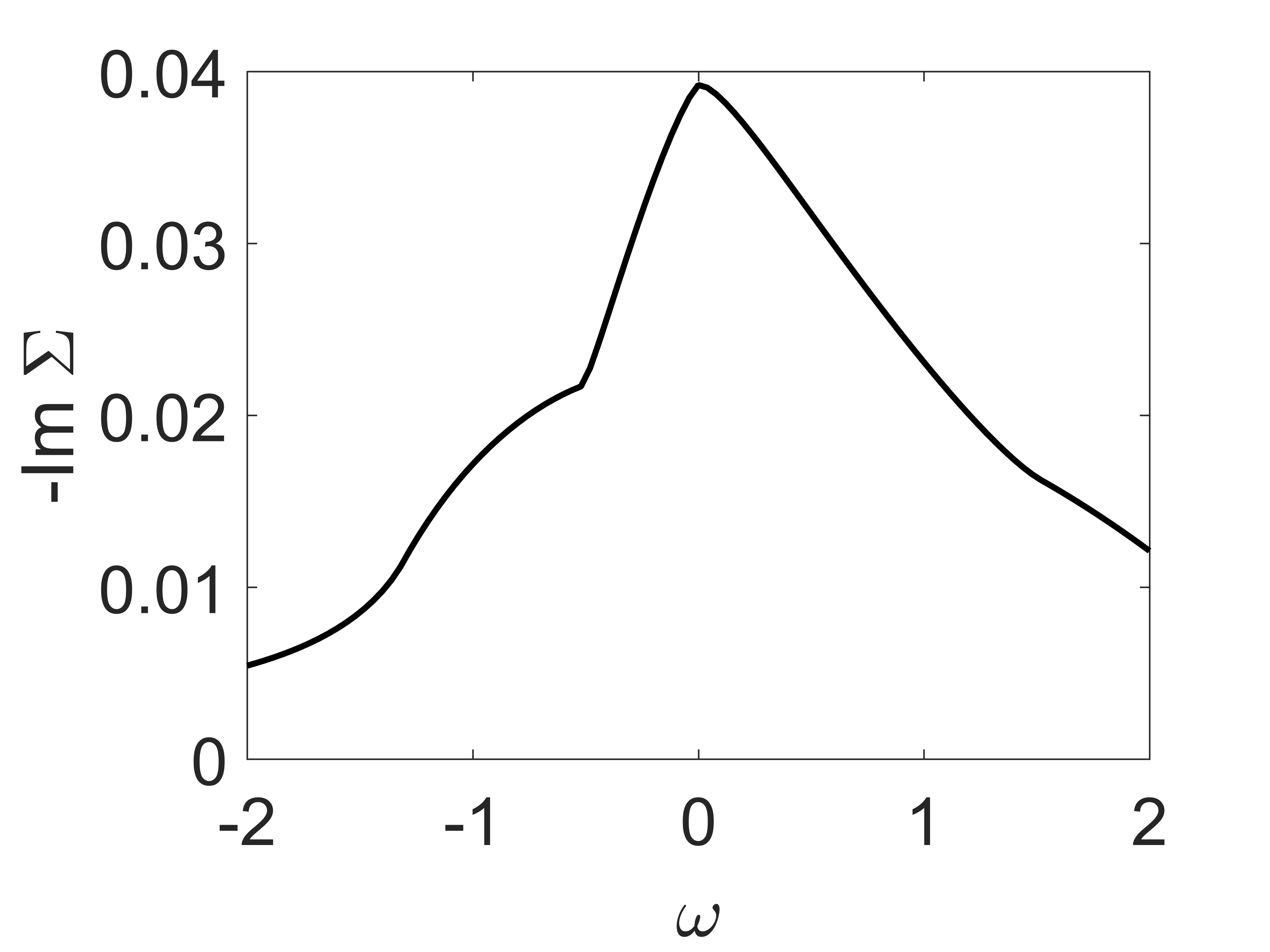}}
    \caption{Plots of the imaginary part of the self-energy as a function of frequency in the case $\alpha = 0.7$ and $d=3$.} 
\end{figure}

\begin{figure}[h]
	\centering 
	\includegraphics[scale=0.5]{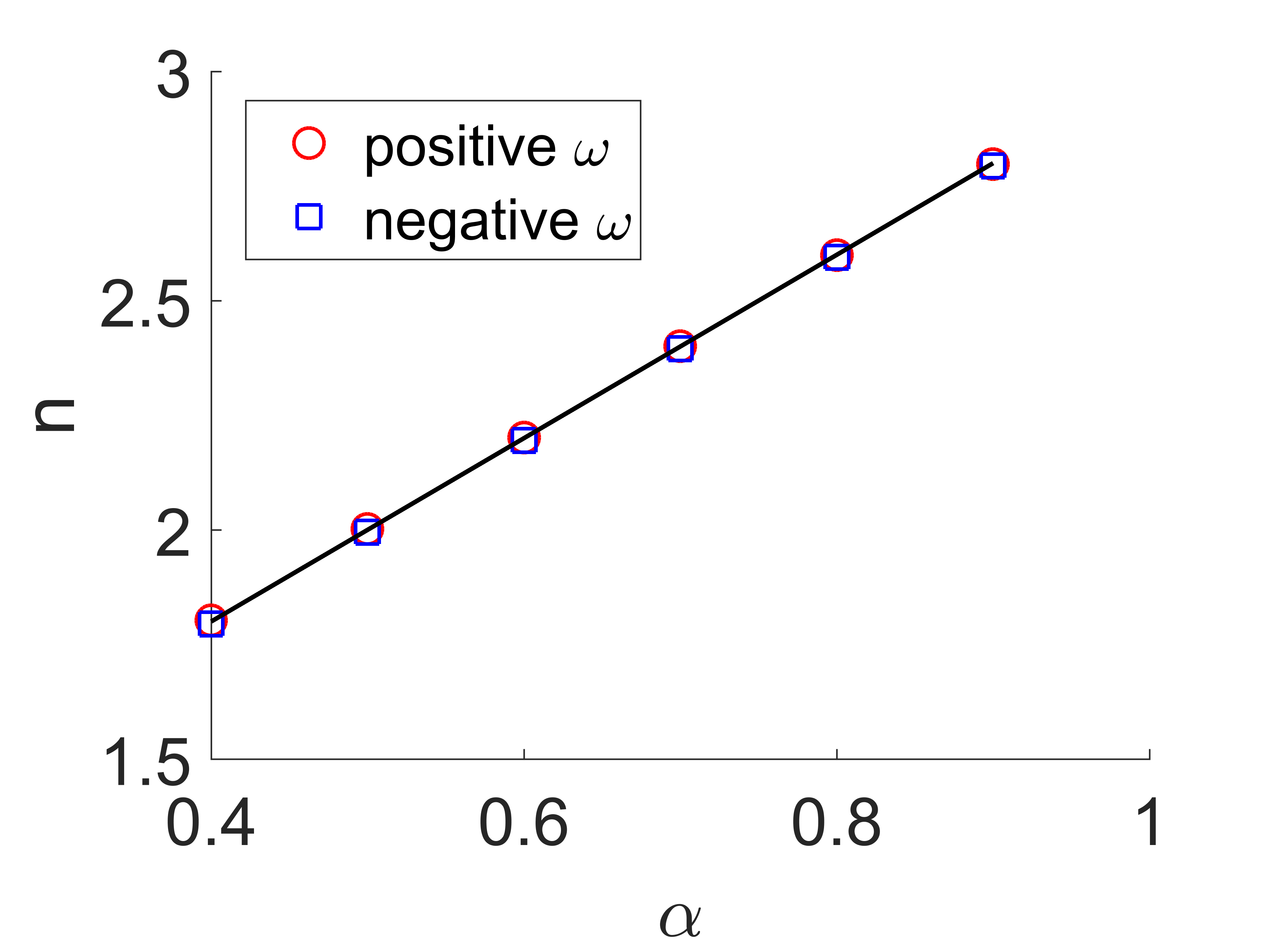}
    \caption{Plot of $\mathrm{Im}\Sigma$'s frequency power law exponent $n$ vs. $\alpha$ at low frequency and temperature ($T = 0.01m$). The spatial dimension in this case is $d = 3$. Squares (Circles) correspond to exponents obtained by fitting the positive (negative) $\omega$ part of $\mathrm{Im}\Sigma$ to a power law. The black line is the plot of $n = d-2+2\alpha$.} \label{fig:lowwpower}
\end{figure}

We numerically study $\mathrm{Im}\Sigma(\omega)$ using the sum of Eq. \ref{eq:im_sigma_b_heav} and the imaginary part of Eq. \ref{eq:sigma_f}. The term $\Sigma_\epsilon$ (Appendix \ref{app:sigma_epsilon}) is now included because the integral over $z$ in Eq. \ref{eq:im_sigma_b_heav} needs a finite lower limit, $\epsilon$. We work in $d = 3$ and use the same parameters as in Sect. \ref{sec:imsig_t}, i.e. $\Lambda = m$, $q = q_f = \sqrt{2}m$, and $v = 0.4$. The results are displayed in Fig. \ref{fig:imsig_w_lowT} for the low-temperature case, and in Fig. \ref{fig:imsig_w_largeT} for the high-temperature case. We find that, in the low-temperature case, $\mathrm{Im}\Sigma(\omega)$ exhibits a power law at low frequencies. This power law has the form 
\beq
\mathrm{Im}\Sigma \sim |\omega|^{d-2+2\alpha}
\eeq
as shown in Fig. \ref{fig:lowwpower}.

\subsection{Discussion}

We found that, at high temperatures, the imaginary part of the electron's self-energy depends linearly on temperature. The linear $T$ behavior is a common feature in a system of fermions interacting with bosons. One well-known example is an electron-phonon system in metals \cite{Mahan2000}. In the context of unparticles, this result is somewhat of a surprise as there is no concept of quantization. Mathematically, the origin of the $T$-linear behavior seems to arise from the summation over the bosonic Matsubara frequency in a self-energy diagram (Fig. \ref{fig:feynman}) which yields the Bose factor $g_B = \frac{1}{2}\coth(\frac{\varepsilon}{2T})$. Since the leading term in the large $T$ expansion of $g_B$ is proportional of $T$, the imaginary part of the self-energy is also linear in $T$. 

At low temperatures, $\mathrm{Im}\Sigma$ of the electron on the Fermi surface exhibits a fractional power law of the form $T^{d-2+2\alpha}$ and $|\omega|^{d-2+2\alpha}$, in qualitative agreement with the experiments \cite{dessau}.  This power law behavior occurs because the excitation energy of electrons close to the Fermi surface becomes linear in momentum i.e. $\varepsilon_{\vec p + \vec q_f} \propto p$. One can see this by noting that when the momentum cutoff $\Lambda$ of the unparticles is much smaller than the Fermi momentum, it is reasonable to drop the ${p}^2/2m$ term in $\varepsilon_{\vec p + \vec q_f}$. For the $d = 3$ case, we give a precise argument that justifies the omission of the ${p}^2$ term in Appendix \ref{app:scaling_imsigma}. The scaling obtained for $\mathrm{Im}\Sigma$ by using $\varepsilon_{\vec p+ \vec q_f}\propto p$ is $\propto T^{1+2\alpha}$ and the error from neglecting the $p^2$ term in $\varepsilon_{\vec p + \vec q_f}$ is $O(T^{2+2\alpha})$. As a result, the error is much smaller than $\mathrm{Im}\Sigma$ in the low-temperature limit. Hence, $\varepsilon_{\vec p + \vec q_f}$ is linear in $p$. We are not able to use the same argument to show the analogous result for $d=2$, but the direct numerical integration reveals that $\mathrm{Im}\Sigma \propto T^{2\alpha}$. This indicates that for the $d=2$ case, $\varepsilon_{\vec p + \vec q_f}$ is also linear in $p$. One thing to note is that the fractional power law comes directly from the anomalous scale, $\alpha$, in the unparticle propagator. The presence of the two branch cuts does not play a major role in determining the low-temperature power law of $\mathrm{Im}\Sigma$. This gives us a hint that, to obtain a power law, we can consider a model in which the anomalous scale appears in the coupling constant i.e. $u\propto p^{\alpha}$ and the bosonic unparticle is replaced by a gapless bosonic particle.

\section{Sum Rules} \label{sec:sumrules}

In this section, we study the effect of the fractional dimension in the propagator on the widely used sum rules. The simplest and most common of these is the spectral or the density of states (DOS) sum rule involving the imaginary part of the electron Green function with excitation energy $\omega_0$ and lifetime $\tau$ given as
\begin{equation}\label{DOSSumRule}
-\int_{-W}^{W} \mathrm{Im} \left( \frac{1}{\omega - \omega_0 + i \tau^{-1}}\right) d\omega = 2\arctan\left( W \tau\right),
\end{equation}
where $W$ is the bandwidth of the electrons in the Fermi sea. In the limit when the excitations are well defined (i.e. when $W\tau \rightarrow \infty$), we obtain the DOS sum rule where the above integral in Eq. \ref{DOSSumRule} sums to $\pi$. The DOS sum rule basically states that the quasiparticle spectral weight, measured, for example, in photoemission or inverse photoemission experiments, is conserved when integrated over all energy scales. To examine the effect of a fractional energy denominator has on the DOS sum rule, we make the substitution
\begin{equation}
 \frac{1}{\omega - \omega_0 + i \tau^{-1}} \rightarrow  \frac{1}{(\omega - \omega_0 + i \tau^{-1} )^{1-\alpha}}.
\end{equation}
The new integral, $I$, we wish to evaluate takes the form
\begin{equation}
I = -\int_{-W}^{W} Im \left[ \frac{\tau^{1-\alpha}}{(\omega \tau - \omega_0 \tau + i )^{1-\alpha}}\right] d\omega.
\end{equation}
One could, in principle, be more general by adding a momentum dependence in the place of a constant excitation energy, which is usually the case. However, it is sufficient to assume it to be a constant in order to extract the singular behavior in $W\tau$ that we anticipate due to the fractional dimension. If we define $\omega \tau = t$,  $\omega_0 \tau = t_0$ and $W\tau = \kappa$ and take the limit $\omega_0\ll W$, we find that
\begin{equation}
I = -\int_{-\kappa}^{\kappa} Im \left[ \frac{\tau^{-\alpha}}{(t + i )^{1-\alpha}}\right] dt.
\end{equation}
To see the effect of a nonzero but small $\alpha$, we can perform a Taylor expansion of the propagator about $\alpha=0$ to obtain
\begin{equation}
\frac{1}{(t+ i)^{1-\alpha}} \approx \frac{1}{i+t} + \alpha \frac{\log(i+t)}{i+t} + O(\alpha^2).
\end{equation}
\begin{figure}[h!] 
\includegraphics[width=1.7in,height=1.3in]{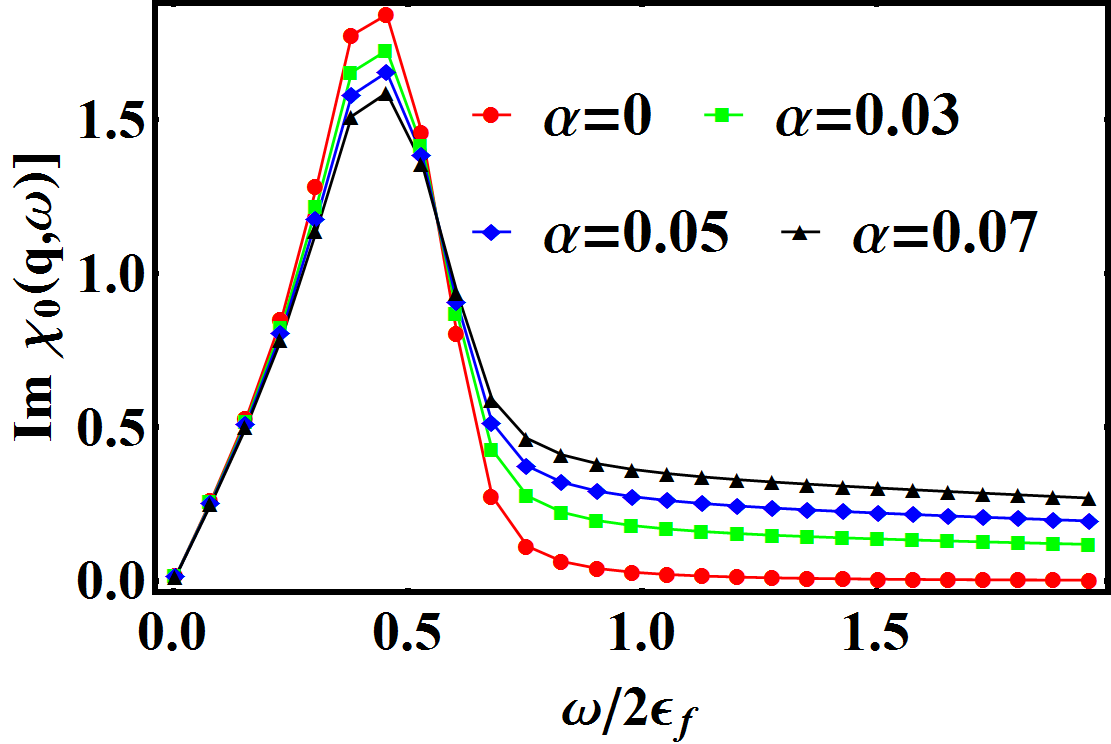}\hfill 
  \includegraphics[width=1.7in,height=1.3in]{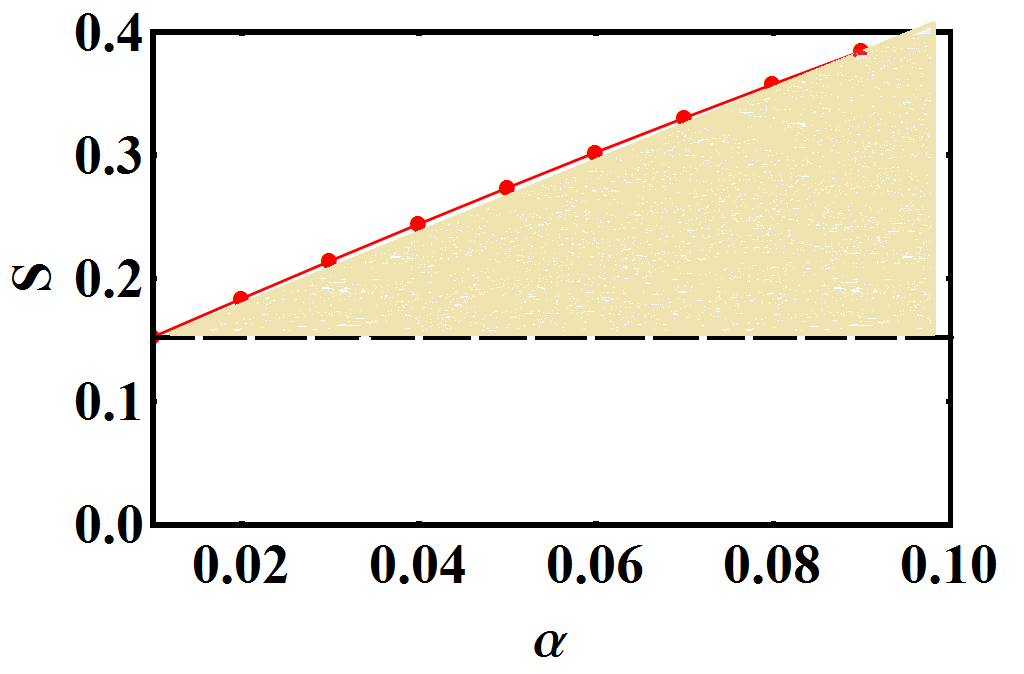}%
\caption{(Left) Plot of the imaginary part of the fermionic density-density correlation function (Lindhard type) as a function of the transferred frequency $\omega$ for various values of $\alpha$. The plots are shown for $ |\vec q| = 0.5 |\vec k_f| $. (Right) Violation of the sum rule integral $S$ as a function of the anomalous dimension $\alpha$. The dashed line shows the value $\alpha=0$ value ($n q^2/m$), the shaded area shows the deviation from this number and the red dots are numerical data. The energy cut-off is fixed at $\frac{\omega_c}{2 \epsilon_f} = 2$} \label{SharkFin}
\end{figure}
The first and second terms are the ``coherent'' and ``incoherent'' contributions to the Green function, respectively. The decoherence in the problem due to a branch cut is now completely transferred into a logarithm in the second term. As a result of this separation, any measurable quantity or correlation function will have contributions from pure coherent and incoherent terms, as well as mixed contributions that come from cross terms. As expected and as will be seen, any violation to the DOS or f-sum rule must come from the incoherent part of the Green function. To finish the evaluation of the integral $I$, we need the imaginary part of the Green function which is given by
\begin{eqnarray}
\mathrm{Im} \left[\frac{1}{(t+ i)^{1-\alpha}}\right] &\approx& -\frac{1}{1+t^2}\\ \nonumber
&& +  \alpha \left[ t \frac{\arg(i+ t)}{1+t^2} - \frac{\log(1+t^2)}{2(1+t^2)} \right] \\ \nonumber
&&+ O(\alpha^2).
\end{eqnarray}
The $t$ integral can now be performed to obtain
\begin{equation}
\frac{-1}{\pi} \int_{-\kappa}^{\kappa} \mathrm{Im}\left[ \frac{\tau^{-\alpha}}{ (i+t)^{1-\alpha}} \right] dt \approx \tau^{-\alpha} \left( 1+ \alpha \log \kappa\right) + O(\alpha^2).
\end{equation}
\begin{figure}[h!]
\includegraphics[height=1.8in]{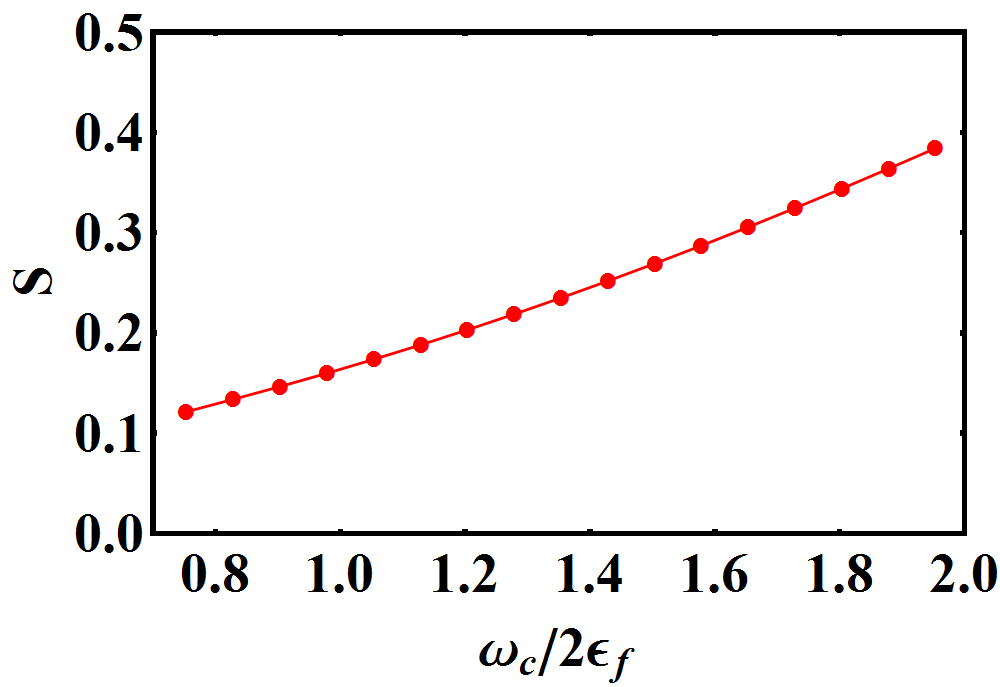}\hfill
\caption{Dependence of the sum rule integral $S$ on the energy cut-off $\frac{\omega_c}{2 \epsilon_f}$. The value of the anomalous dimension parameter is fixed at $\alpha = 0.09$.}\label{SumRule-Violate-Cutoff}
\end{figure}
Thus, in the limit of a sharp quasiparticle peak and small $\alpha$, the correction due to a fractional energy dimension in the Green function is proportional to $\alpha$ and diverges logarithmically. 

We will now numerically evaluate the sum rule violation for the density-density correlation function (f-sum rule) which is traditionally given as
\begin{equation}\label{fsum}
S \equiv \frac{-2}{\pi} \int_0^{\infty} \omega d\omega \chi''_0(\omega,\vec q) = \frac{n q^2}{m},
\end{equation}
where $n$ is the electron density, $q$ is the absolute value of $\vec q$, and $m$ is the electron mass. Physically, Eq. \ref{fsum} says that the total electron density contributing to the density response in a certain energy window is given by the area under the curve of the experimentally measured density-density correlation function in that energy window. Similar sum rules can be formulated for the optical conductivity or the dielectric response where one can equivalently estimate the charge density. Such a counting procedure of the particle or charge density forms a consequence of the single particle description of a response system. However, there is no reason to expect that such a counting procedure should continue to hold in the presence of interactions. To test this conjecture, we artificially introduce a fractional dimension to the fermionic Matsubara Green function
\begin{equation}
 \frac{1}{i \omega_n - \epsilon(\vec k)} \rightarrow  \frac{1}{( i \omega_n - \epsilon(\vec k) )^{1-\alpha}},
\end{equation}
where $\omega_n$ are the fermionic Matsubara frequencies, $\epsilon(\vec k)$ is the electron band energy with a Fermi energy $\epsilon_f$ (and momentum $\vec k_f$), and $\alpha$ is the fractional dimension. Although the form of the above substitution is not strictly that of a scale-invariant fermionic unparticle, it gives us a flavor of the sum rule violation. The breakdown of the f-sum rule for different values of the fractional dimension $\alpha$ is shown numerically in Fig. \ref{SharkFin}. The left panel shows the characteristic shark-fin shaped Lindhard response obtained for $q/k_f = 0.5$ and different values of $\alpha$. Clearly, there is a high energy tail that develops with nonzero values of $\alpha$. The right panel shows the sum $S$ with the energy cutoff fixed at $\omega_c = 4 \epsilon_f$ and $|\vec q| = 0.5 |\vec k_f|$ with the dashed line showing the value evaluated at $\alpha = 0$. The shaded region quantifies the deviation from the $\alpha=0$ value with the deviation being linearly proportional to $\alpha$, just like in the case for the DOS scenario. The sum rule violation due to the cutoff dependence (see Fig. \ref{SumRule-Violate-Cutoff}), on the other hand, deviates faster than the DOS case where the dependence on the energy cutoff was logarithmic.\\
\newline

\section{Conclusion} \label{sec:conclusion}
We have studied the interaction of a scale-invariant sector with electron quasiparticles, and found that it is the bosonic character of  scalar unparticles that gives rise to the linear $T$ in $\mathrm{Im}\Sigma$ at high temperatures. At low temperatures, the electrons on the Fermi surface become scale-covariant with $z=1$. One can then simply use the scaling analysis to show that $\mathrm{Im}\Sigma \propto T^{d-2+2\alpha}$. Similar results hold also for the frequency dependence, as indicative of the power-law liquid seen experimentally \cite{dessau}.   It would be interesting to see how the result we find here translate into a temperature dependence of an electrical resistivity, $\rho$, which is proportional to the relaxation time, $\tau_{rl}$. 

The logarithmic divergence in the spectral sum rule is not unexpected. The long high-energy tails, acquired as a result of the anomalous dimension, go to infinity, giving rise to divergent integrals if a high energy cutoff is not imposed. However, in order to recover the sum rule, one may need to define a ``fractional'' energy integral which absorbs or cancels the logarithmic term. This seems like the most natural prescription to derive a useful sum rule as the necessary route to obtaining a fractional dimension in the Green function involves fractional calculus \cite{Herrmann2014,Kilbas2006,Millercalc,Samko1993}.  

Because mass is energy, integrating over mass is equivalent to integrating over all energy scales.  In doped Mott insulators, removing a single hole \cite{phillipsrmp,sawatzky} leads to spectral weight transfer over all energy scales.  This gives rise to an incoherent background in the electron spectral function.  Unparticles are an attempt to model such incoherence, and the continuous mass formalism is designed to capture this aspect of Mott physics.  That unparticles effectively give rise to power-law contributions to the electron self-energy points to a possible physical mechanism underlying power-law liquids \cite{dessau}.\\

\noindent \textbf{Acknowledgements} We thank the NSF DMR-1461952 for partial funding of this project. KL is supported by the Department of Physics at the University of Illinois and a scholarship from the Ministry of Science and Technology, Royal Thai Government. ZL is supported by a scholarship from the Agency of Science, Technology and Research. CS and PWP are supported by the Center for Emergent Superconductivity, a DOE Energy Frontier Research Center, Grant No. DE-AC0298CH1088. 

\onecolumngrid 

\appendix

\section{Calculation of the Electron-Unparticle Self Energy} \label{app:matsum}
\begin{figure}[h]
	\includegraphics[scale=0.45]{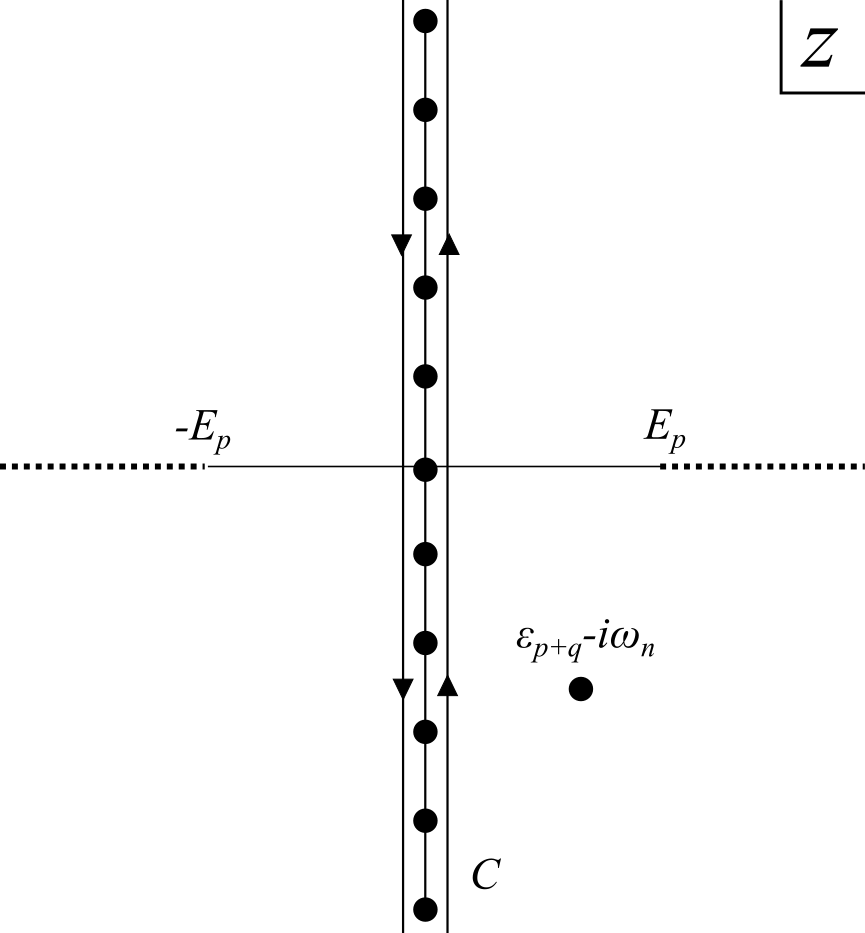}
    \caption{The contour $C$ of the integral over $z$ in Eq. \ref{eq:sigma_con}. A solid dot represents a first order pole. A dotted line represents a branch cut.}
    \label{fig:contour}
\end{figure}

The summation over $\omega_n$ in Eq. \ref{eq:sigma} can be converted into a contour integral:
\beq \label{eq:sigma_con}
\Sigma_n(\vec q) = u^2\int \frac{d^dp}{(2\pi)^d}\int\limits_{C} \frac{dz}{2\pi i} \frac{g_B(z)}{(E^2_{\vec p}-z^2)^{1-\alpha}(z+i\omega_n-\varepsilon_{\vec p + \vec q})}
\eeq
where $g_B(z) = \frac{1}{2}\coth\frac{\beta z}{2}$ is a bosonic pole function and the contour $C$ is shown in Fig. \ref{fig:contour}.

It is tempting to rewrite the term $\frac{1}{(E^2_{\vec p}-z^2)^{1-\alpha}}$ as $\frac{e^{i\pi(1-\alpha)}}{(z-E_{\vec p})^{1-\alpha}(z+E_{\vec p})^{1-\alpha}}$. To do this properly, we need to choose a proper Riemann surface. Such choices must satisfy the condition that upon performing the residue integral over the poles along the imaginary axis the integral $\int\limits_C dz\frac{e^{i\pi(1-\alpha)}}{(z-E_{\vec p})^{1-\alpha}(z+E_{\vec p})^{1-\alpha}}$ is real and is equal to $T\sum\limits_{m}\frac{1}{(\omega_m^2+E^2_{\vec p})^{1-\alpha}}$. The following choice of Riemann surface satisfies the above condition. The base of the term $(z-E_{\vec p})^{1-\alpha}$ is chosen to have its phase angle in the range $0 \le \theta_1 < 2\pi$ whereas the base of the term $(z+E_{\vec p})^{1-\alpha}$ is chosen to have its phase angle in the range $-\pi \le \theta_2 < \pi$. Performing residue integrals along the imaginary axis in Eq. \ref{eq:sigma_con} leads to substituting $z=i\omega_m$ into the terms $(z-E_{\vec p})^{1-\alpha}$ and $(z+E_{\vec p})^{1-\alpha}$. The results are
\beq
(z+E_{\vec p})^{1-\alpha}\bigg|_{z = i\omega_m} &=& |i\omega_m - E_{\vec p}|^{1-\alpha} \exp(i(\pi-\arctan(\frac{\omega_m}{E_{\vec p}}))(1-\alpha)) \nonumber \\
(z-E_{\vec p})^{1-\alpha}\bigg|_{z = i\omega_m} &=& |i\omega_m + E_{\vec p}|^{1-\alpha} \exp(i\arctan(\frac{\omega_m}{E_{\vec p}})(1-\alpha)). \nonumber
\eeq
Consequently, $\frac{e^{i\pi(1-\alpha)}}{(z-E_{\vec p})^{1-\alpha}(z+E_{\vec p})^{1-\alpha}}\bigg|_{z=i\omega_m} = \frac{1}{(\omega_m^2 + E^2_{\vec p})^{1-\alpha}}$ and so we can rewrite $\frac{1}{(E^2_{\vec p}-z^2)^{1-\alpha}} = \frac{e^{i\pi(1-\alpha)}}{(z-E_{\vec p})^{1-\alpha}(z+E_{\vec p})^{1-\alpha}}$ in Eq. \ref{eq:sigma_con}. By deforming the contour to a circle of large radius $R$, we find that there are four contributions to the self energy:
\beq
\Sigma = \Sigma_{F} + \Sigma_{B} + \Sigma_{\epsilon} + \Sigma_{R}.
\eeq
$\Sigma_{F}$ comes from the pole at $z = \varepsilon_{\vec p + \vec q} - i\omega_n$. $\Sigma_{B}$ comes from the two branch cuts. $\Sigma_{\epsilon}$ comes from the small circles of radius $\epsilon$ around the two branch points at $\pm E_{\vec p}$. Finally, $\Sigma_{R}$ comes form the large circle of radius $R$. We discuss these four terms below.

\subsection{$\Sigma_{F}$} \label{app:sigma_f}
Calculating the residue of Eq. \ref{eq:sigma_con} at $z = \varepsilon_{\vec p + \vec q} - i\omega_n$, we obtain
\beq
\Sigma_{F,n}(\vec q) = -e^{i\pi(1-\alpha)}u^2\int \frac{d^dp}{(2\pi)^d} \frac{g_F(\varepsilon_{\vec p + \vec q})}{(\varepsilon_{\vec p + \vec q}-E_{\vec p}-i\omega_n)^{1-\alpha}(\varepsilon_{\vec p + \vec q}+E_{\vec p}-i\omega_n)^{1-\alpha}}
\eeq
Here we use
\beq
g_B(z+i\omega_n) = \frac{1}{2}\coth\frac{\beta (z+i\frac{(2n-1)\pi}{\beta})}{2} = \frac{1}{2}\tanh \frac{\beta z}{2} = g_F(z).
\eeq
to simplify the result. Note that the phase angle of the term $\varepsilon_{\vec p + \vec q}-E_{\vec p}-i\omega_n$ when raise to the power $1-\alpha$ is defined in the range $0\le\theta_1<2\pi$. We can covert the phase angle to be $-\pi \le \theta_2 < \pi$ by $(\varepsilon_{\vec p + \vec q}-E_{\vec p}-i\omega_n)^{1-\alpha}\big|_{\theta_1} \rightarrow e^{i\pi(1-\alpha)}[e^{-i\pi}(\varepsilon_{\vec p + \vec q}-E_{\vec p}-i\omega_n)]^{1-\alpha}\big|_{\theta_2}$. Here $x^{1-\alpha}\big|_{\theta_i}$ means computing $x^{1-\alpha}$ with the definition of $\theta_i$. Changing the definition to $\theta_2$, $\Sigma_F$ becomes
\beq \label{eq:sigma_f}
\Sigma_{F,n}(\vec q) = -u^2\int \frac{d^dp}{(2\pi)^d} \frac{g_F(\varepsilon_{\vec p + \vec q})}{(-\varepsilon_{\vec p + \vec q}+E_{\vec p}+i\omega_n)^{1-\alpha}(\varepsilon_{\vec p + \vec q}+E_{\vec p}-i\omega_n)^{1-\alpha}}.
\eeq

\subsection{$\Sigma_{B}$} \label{app:sigma_b}
We can rewrite this contribution to the self energy as discontinuities across the two branch cuts as
\beq
\Sigma_{B,n}(\vec q) &=& e^{i\pi(1-\alpha)}u^2\int \frac{d^dp}{(2\pi)^d}\int\limits_{\epsilon+E_{\vec p}}^{\infty} \frac{dz}{2\pi i} \frac{g_B(z)}{(z+i\omega_n-\epsilon_{\vec p + \vec q})}\frac{1}{(z+E_{\vec p})^{1-\alpha}}\bigg(\frac{1}{(z^{+}-E_{\vec p})^{1-\alpha}} - \frac{1}{(z^{-}-E_{\vec p})^{1-\alpha}}\bigg) \nonumber \\
&& + e^{i\pi(1-\alpha)}u^2\int \frac{d^dp}{(2\pi)^d}\int\limits_{-\infty}^{-\epsilon-E_{\vec p}} \frac{dz}{2\pi i} \frac{g_B(z)}{(z+i\omega_n-\epsilon_{\vec p + \vec q})}\frac{1}{(z-E_{\vec p})^{1-\alpha}}\bigg(\frac{1}{(z^{+}+E_{\vec p})^{1-\alpha}} - \frac{1}{(z^{-}+E_{\vec p})^{1-\alpha}}\bigg) \nonumber \\
\eeq
where $z^{\pm} \equiv z \pm i\eta$. Using definitions of the phase angles, $\theta_1$ and $\theta_2$, we define above, the discontinuities are given by
\beq
\frac{1}{(z^{+}-E_{\vec p})^{1-\alpha}} - \frac{1}{(z^{-}-E_{\vec p}){1-\alpha}} &=& \frac{2i\sin(\pi\alpha) e^{-i\pi(1-\alpha)}}{|z-E_{\vec p}|^{1-\alpha}} \nonumber \\
\frac{1}{(z^{+}+E_{\vec p})^{1-\alpha}} - \frac{1}{(z^{-}+E_{\vec p}){1-\alpha}} &=& -\frac{2i\sin(\pi\alpha)}{|z+E_{\vec p}|^{1-\alpha}}. \nonumber
\eeq
Substituting them into $\Sigma_B$, we obtain
\beq
\Sigma_{B,n}(\vec q) = \frac{\sin(\pi\alpha)}{\pi}u^2\int\frac{d^dp}{(2\pi)^d}\int\limits_{\epsilon+E_{\vec p}}^{\infty}dz \frac{1}{(z+E_{\vec p})^{1-\alpha}(z-E_{\vec p})^{1-\alpha}}\bigg(\frac{g_B(z)}{(z+i\omega_n-\varepsilon_{\vec p + \vec q})} + \frac{g_B(-z)}{(z-i\omega_n+\varepsilon_{\vec p + \vec q})}\bigg).
\eeq

\subsection{$\Sigma_\epsilon$} \label{app:sigma_epsilon}
For the integral around the branch point at $z = E_{\vec p}$, we let $z = E_{\vec p} + \epsilon e^{i\theta}$ with $0<\theta<2\pi$ and for the integral around the branch point at $z = -E_{\vec p}$, we let $z = -E_{\vec p} + \epsilon e^{i\theta}$ with $-\pi<\theta<\pi$. The result is
\beq
\Sigma_{\epsilon,n}(\vec q) &=& -\epsilon^{\alpha} e^{i\pi(1-\alpha)}u^2\int\frac{d^dp}{(2\pi)^d}\int\limits_{0}^{2\pi}\frac{d\theta}{2\pi} \frac{ e^{i\alpha\theta}g_B(E_{\vec p}+\epsilon e^{i\theta})}{ (2E_{\vec p}+\epsilon e^{i\theta})^{1-\alpha}(E_{\vec p}+i\omega_n-\varepsilon_{\vec p + \vec q}+\epsilon e^{i\theta})} \nonumber \\
&& - \epsilon^{\alpha} e^{i\pi(1-\alpha)}u^2\int\frac{d^dp}{(2\pi)^d}\int\limits_{-\pi}^{\pi}\frac{d\theta}{2\pi} \frac{ e^{i\alpha\theta}g_B(-E_{\vec p}+\epsilon e^{i\theta})}{(-2E_{\vec p}+\epsilon e^{i\theta})^{1-\alpha}(-E_{\vec p}+i\omega_n-\varepsilon_{\vec p + \vec q}+\epsilon e^{i\theta})}.
\eeq
This means $\lim\limits_{\epsilon\rightarrow0} \Sigma_{\epsilon,n}(\vec q) = 0$ when $\alpha>0$.

\subsection{$\Sigma_R$} \label{app:sigma_R}
For the contribution from the large circle at radius $R$, we let $z = R e^{i\theta}$. One has
\beq
\Sigma_{R,n}(\vec q) &=& e^{i\pi(1-\alpha)}u^2\int \frac{d^dp}{(2\pi)^d}\int\limits_{0}^{2\pi}\frac{d\theta}{2\pi} \frac{Re^{i\theta}g_B(Re^{i\theta})}{(Re^{i\theta}+E_{\vec p})^{1-\alpha}(Re^{i\theta}-E_{\vec p})^{1-\alpha}(Re^{i\theta}+i\omega_n-\varepsilon_{\vec p + \vec q})}. \nonumber
\eeq
In the limit of large $R$, we have
\beq
|\Sigma_{R,n}(\vec q)| &\approx&  R^ {-2(1-\alpha)}u^2 \int\frac{d^dp}{(2\pi)^d}\int\limits_{0}^{2\pi}\frac{d\theta}{2\pi}|g_B(Re^{i\theta})|ใ
\eeq
This means $\lim\limits_{R\rightarrow\infty}\Sigma_{R,n}(\vec q) = 0$ when $\alpha<1$.

\subsection{Total $\Sigma$}
By restricting the exponent $\alpha$ to be in the range $0<\alpha<1$, the terms $\Sigma_{\epsilon}$ and $\Sigma_R$ can be omitted. Combining $\Sigma_B$ and $\Sigma_F$ from subsections \ref{app:sigma_f} and \ref{app:sigma_b} and then performing analytic continuation $i\omega_n \rightarrow \omega+i\eta$, one obtains the result
\beq
\Sigma(\omega,\vec q) &=& -u^2 \int \frac{d^dp}{(2\pi)^d} \frac{g_F(\varepsilon_{\vec p + \vec q})}{(-\varepsilon_{\vec p + \vec q}+E_{\vec p}+\omega+i\eta)^{1-\alpha}(\varepsilon_{\vec p + \vec q}+E_{\vec p}-\omega-i\eta)^{1-\alpha}} \nonumber \\
&& + \frac{\sin(\pi\alpha)}{\pi}u^2 \int\frac{d^dp}{(2\pi)^d}\int\limits_{E_{\vec p}}^{\infty}dz \frac{g_B(z)}{(z+E_{\vec p})^{1-\alpha}(z-E_{\vec p})^{1-\alpha}}\bigg(\frac{1}{(z+\omega+i\eta-\varepsilon_{\vec p + \vec q})} - \frac{1}{(z-\omega-i\eta+\varepsilon_{\vec p + \vec q})}\bigg). \nonumber \\ \label{eq:sigma_total}
\eeq

\section{Scaling of the Imaginary Part of the Self Energy at Low Temperature} \label{app:scaling_imsigma}
In this section, we analyze Eq. \ref{eq:sigma_total} at low temperature. We work with electrons on the Fermi surface ($\vec q = \vec q_f$) and $\omega = 0$ in $d = 3$ spatial dimensions.

\subsection{$\Sigma_B$}
We start by considering $\Sigma_B$ (the second term of Eq. \ref{eq:sigma_total}). Using the identity $\frac{1}{x\pm i\eta} = P(\frac{1}{x}) \mp i\pi\delta(x)$ and taking the imaginary part yields
\beq
\mathrm{Im} \ \Sigma_B(\omega,\vec q) &=& -\frac{\sin(\pi\alpha)}{4\pi^2}u^2\int\limits_0^{\Lambda}dp\int\limits_{-1}^{1}dx  \int\limits_{E_{\vec p}}^{\infty}dz  \frac{p^{2}g_B(z)}{(z+E_{\vec p})^{1-\alpha}(z-E_{\vec p})^{1-\alpha}}\bigg(\delta(z+\omega-\frac{p^2}{2m}-\frac{pqx}{m}-\frac{q^2}{2m}+\varepsilon_f) \nonumber \\
&& \tabiv \ \ \ \ \  + \delta(z-\omega+\frac{p^2}{2m}+\frac{pqx}{m}+\frac{q^2}{2m}+\varepsilon_f)\bigg)
\eeq
where $P$ denotes the principal part of the Cauchy principal integral. Here we let $x = \cos\theta$. Since the range of $x$ is from $-1$ to $1$, the integral of the two delta functions over $x$ yields two Heaviside functions,
\beq
\frac{m}{pq}\Theta(1-\frac{m}{pq}|z - \frac{p^2}{2m} - \frac{q^2}{2m} + \varepsilon_f +\omega|) + \frac{m}{pq}\Theta(1-\frac{m}{pq}|z + \frac{p^2}{2m} + \frac{q^2}{2m} - \varepsilon_f -\omega|).
\eeq
They put restrictions on the range of the integral over $z$. The imaginary part of $\Sigma_B$ is now
\begin{align} \label{eq:im_sigma_b_heav} 
\mathrm{Im} \ \Sigma_B(\omega,\vec q) = -\frac{\sin(\pi\alpha)}{4\pi^2}\frac{mu^2}{q}\int\limits_0^{\Lambda}dp  \int\limits_{E_{\vec p}}^{\infty}dz  \frac{p g_B(z)}{(z+E_{\vec p})^{1-\alpha}(z-E_{\vec p})^{1-\alpha}}\bigg(& \Theta(1-\frac{m}{pq}|z - \frac{p^2}{2m} - \frac{q^2}{2m} + \varepsilon_f +\omega|) \nonumber \\
& + \Theta(1-\frac{m}{pq}|z + \frac{p^2}{2m} + \frac{q^2}{2m} - \varepsilon_f -\omega|)\bigg).
\end{align}
We substitute $E_{\vec p} = pv$, $\vec q = \vec q_f$, and $\omega = 0$ into the above equation, and then perform a change of variable $z\rightarrow p(z+v)$. The result is
\begin{align}
\mathrm{Im} \ \Sigma_B(\omega=0,\vec q_f) = -\frac{\sin(\pi\alpha)}{4\pi^2}\frac{mu^2}{q_f}\int\limits_0^{\Lambda}dp  \int\limits_{0}^{\infty}dz  \frac{p^{2\alpha}g_B(p(z+v))}{(z+2v)^{1-\alpha}z^{1-\alpha}}\bigg(& \Theta(q_f-m|z + v - \frac{p}{2m} |) \nonumber \\
& + \Theta(q_f-m|z+v + \frac{p}{2m} |)\bigg) 
\end{align}
In the case of small momentum cutoff $\frac{\Lambda}{2m} < \frac{q_f}{m} - v$ and small velocity $v<\frac{q_f}{m}$, the range of the $z$-integral is given by
\beq
\int\limits_{0}^{\infty} dz \bigg(\Theta(q_f-m|z + v - \frac{p}{2m} |) + \Theta(q_f-m|z+v + \frac{p}{2m} |)\bigg) \longrightarrow \int\limits_{0}^{\frac{1}{m}(q_f-mv+\frac{p}{2})} dz + \int\limits_{0}^{\frac{1}{m}(q_f-mv-\frac{p}{2})} dz.
\eeq
Thus, the imaginary part of $\Sigma_B$ is
\beq
\mathrm{Im} \ \Sigma_B(\omega=0,\vec q_f) &=&-\frac{\sin(\pi\alpha)}{4\pi^2}\frac{mu^2}{q_f}\int\limits_0^{\Lambda}dp   \bigg( \int\limits_{0}^{\frac{1}{m}(q_f-mv+\frac{p}{2})} dz + \int\limits_{0}^{\frac{1}{m}(q_f-mv-\frac{p}{2})} dz \bigg) \frac{p^{2\alpha}g_B(p(z+v))}{(z+2v)^{1-\alpha} z^{1-\alpha}}.
\eeq

\subsection{$\Sigma_F$}
We now turn to the fermionic part of the self energy (the first term of Eq. \ref{eq:sigma_total}). For $\alpha<1$ and $\eta \rightarrow 0^{+}$, we have
\beq
\frac{1}{(x\pm i\eta)^{1-\alpha}} = \frac{1}{|x|^{1-\alpha}}((\Theta(x)-\Theta(-x)\cos\pi\alpha) \mp i\Theta(-x)\sin\pi\alpha).
\eeq
Applying these identities, one can show that $\mathrm{Im}\Sigma_F$ is given by
\beq
\mathrm{Im} \Sigma_F(\omega,\vec q) = -\sin(\pi\alpha)u^2 && \int\frac{d^dp}{(2\pi)^d} \frac{g_F(\varepsilon_{\vec p + \vec q})}{|-\varepsilon_{\vec p + \vec q}+E_{\vec p}+\omega|^{1-\alpha}|\varepsilon_{\vec p + \vec q}+E_{\vec p}-\omega|^{1-\alpha}} \nonumber \\
&& \times (\Theta(-\varepsilon_{\vec p + \vec q}+E_{\vec p}+\omega)\Theta(-\varepsilon_{\vec p + \vec q}-E_{\vec p}+\omega)-\Theta(\varepsilon_{\vec p + \vec q}+E_{\vec p}-\omega)\Theta(\varepsilon_{\vec p + \vec q}-E_{\vec p}-\omega)). \nonumber \\
\eeq
We set $d = 3$, $q = q_f$, $\omega = 0$, and $E_{\vec p} = pv$. The result is
\beq
\mathrm{Im} \ \Sigma_F(\omega=0,\vec q_f) =&&  -\frac{\sin(\pi\alpha)}{4\pi^2}u^2\int\limits_{0}^{\Lambda}dp \int\limits_{-1}^{1}dx \frac{p^2 g_F(\frac{p^2}{2m}+\frac{pq_f x}{m})}{|-\frac{p^2}{2m}-\frac{pq_f x}{m}+pv|^{1-\alpha}|\frac{p^2}{2m}+\frac{pq_f x}{m}+pv|^{1-\alpha}} \nonumber \\
&& \times (\Theta(-\frac{p^2}{2m}-\frac{pq_f x}{m}+pv)\Theta(-\frac{p^2}{2m}-\frac{pq_f x}{m}-pv)-\Theta(\frac{p^2}{2m}+\frac{pq_f x}{m}+pv)\Theta(\frac{p^2}{2m}+\frac{pq_f x}{m}-pv)). \nonumber \\
\eeq 
The first term restricts $x$ to be $-1<x<\frac{m}{q_f}(-\frac{p}{2m}-v)$. With the assumptions that $v<\frac{q_f}{m}$ and $\frac{\Lambda}{2}<q_f-mv$, we find $\frac{m}{q_f}(-\frac{p}{2m}-v)>-1$ for the whole range of $p$. As a result, the integration limits of the first term become $\int\limits_{0}^{\Lambda}dp\int\limits_{-1}^{-\frac{1}{q_f}(\frac{p}{2}+mv)}dx$. The second term restricts $x$ to be $\frac{m}{q_f}(-\frac{p}{2m}+v)<x<1$. With the assumptions that $v<\frac{q_f}{m}$ and $\frac{\Lambda}{2}<q_f-mv$, we find $-1<\frac{m}{q_f}(-\frac{p}{2m}+v)<1$ for the whole range of $p$. As a result, the integration limits of the second term become $\int\limits_{0}^{\Lambda}dp\int\limits_{\frac{1}{q_f}(-\frac{p}{2}+mv)}^{1}dx$. The imaginary part of $\Sigma_F$ is now
\beq
\mathrm{Im} \ \Sigma_F(\omega=0,\vec q_f) &=&  -\frac{\sin(\pi\alpha)}{4\pi^2}u^2\int\limits_{0}^{\Lambda}dp \bigg(\int\limits_{-1}^{-\frac{1}{q_f}(\frac{p}{2}+mv)}dx - \int\limits_{\frac{1}{q_f}(-\frac{p}{2}+mv)}^{1}dx \bigg) \frac{p^2 g_F(\frac{p^2}{2m}+\frac{pq_f x}{m})}{|-\frac{p^2}{2m}-\frac{pq_f x}{m}+pv|^{1-\alpha}|\frac{p^2}{2m}+\frac{pq_f x}{m}+pv|^{1-\alpha}} \nonumber \\
&=& -\frac{\sin(\pi\alpha)}{4\pi^2}\frac{mu^2}{q_f}\int\limits_{0}^{\Lambda}dp \bigg(\int\limits_{\frac{p}{2m}-\frac{q_f}{m}}^{-v}dx' - \int\limits_{v}^{\frac{p}{2m}+\frac{q_f}{m}}dx' \bigg) \frac{p^{2\alpha} g_F(px')}{|x'-v|^{1-\alpha}|x'+v|^{1-\alpha}} \nonumber \\
&=& -\frac{\sin(\pi\alpha)}{4\pi^2}\frac{mu^2}{q_f}\int\limits_{0}^{\Lambda}dp \bigg(-\int\limits_{v}^{-\frac{p}{2m}+\frac{q_f}{m}}dx' - \int\limits_{v}^{\frac{p}{2m}+\frac{q_f}{m}}dx' \bigg) \frac{p^{2\alpha} g_F(px')}{|x'-v|^{1-\alpha}|x'+v|^{1-\alpha}} \nonumber \\
&=& \frac{\sin(\pi\alpha)}{4\pi^2}\frac{mu^2}{q_f}\int\limits_{0}^{\Lambda}dp \bigg(\int\limits_{0}^{-\frac{p}{2m}+\frac{q_f}{m}-v}dz + \int\limits_{0}^{\frac{p}{2m}+\frac{q_f}{m}-v}dz \bigg) \frac{p^{2\alpha} g_F(p(z+v))}{z^{1-\alpha}(z+2v)^{1-\alpha}}. 
\eeq 
On the first line, we make a change of variable $x' = \frac{p}{2m} + \frac{q_f x}{m}$ and, on the first integral of the second line, we let $x' \rightarrow -x'$. Finally, we make a shift $x' \rightarrow z = x'-v$ on the third line. 

\subsection{Total $\Sigma$}
Combing $\mathrm{Im}\Sigma_F$ and $\mathrm{Im}\Sigma_B$, one has
\beq \label{eq:sigma_combine_final_3d}
\mathrm{Im} \Sigma(\omega=0,\vec q_f) &=&  -\frac{\sin(\pi\alpha)}{4\pi^2}\frac{mu^2}{q_f}\int\limits_{0}^{\Lambda}dp \bigg(\int\limits_{0}^{-\frac{p}{2m}+\frac{q_f}{m}-v}dz + \int\limits_{0}^{\frac{p}{2m}+\frac{q_f}{m}-v}dz \bigg) p^{2\alpha}\frac{g_B(p(z+v))-g_F(p(z+v))}{z^{1-\alpha}(z+2v)^{1-\alpha}} \nonumber \\
&=& -\frac{\sin(\pi\alpha)}{4\pi^2}\frac{mu^2}{q_f}\int\limits_{0}^{\Lambda}dp \bigg(\int\limits_{0}^{-\frac{p}{2m}+\frac{q_f}{m}-v}dz + \int\limits_{0}^{\frac{p}{2m}+\frac{q_f}{m}-v}dz \bigg)  p^{2\alpha}\frac{n_B(p(z+v))+n_F(p(z+v))}{z^{1-\alpha}(z+2v)^{1-\alpha}} 
\eeq
Here we use the identities $g_B(z) = \frac{1}{2} + n_B(z)$ and $g_F(z) = \frac{1}{2} - n_F(z)$.
We split the limits of the integrals as follows:
\beq
\int\limits_{0}^{\frac{1}{m}(q_f-mv+\frac{p}{2})} dz &\longrightarrow& \int\limits_{0}^{\frac{1}{m}(q_f-mv)} + \int\limits_{\frac{1}{m}(q_f-mv)}^{\frac{1}{m}(q_f-mv+\frac{p}{2})} \nonumber \\
\int\limits_{0}^{\frac{1}{m}(q_f-mv-\frac{p}{2})} dz &\longrightarrow& \int\limits_{0}^{\frac{1}{m}(q_f-mv)} + \int\limits_{\frac{1}{m}(q_f-mv)}^{\frac{1}{m}(q_f-mv-\frac{p}{2})}.
\eeq
The imaginary part of $\Sigma$ is now given by
\begin{align} \label{eq:sigmab1}
\mathrm{Im} \ \Sigma(T) =& - \frac{\sin(\pi\alpha)}{2\pi^2}\frac{mu^2}{q_f}\int\limits_0^{\Lambda}dp  \int\limits_{0}^{\frac{1}{m}(q_f-mv)}dz   p^{2\alpha}\frac{n_B(p(z+v))+n_F(p(z+v))}{z^{1-\alpha}(z+2v)^{1-\alpha}} \nonumber \\
& - \frac{\sin(\pi\alpha)}{4\pi^2}\frac{mu^2}{q_f} \int\limits_0^{\Lambda}dp  \bigg( \int\limits_{\frac{1}{m}(q_f-mv)}^{\frac{1}{m}(q_f-mv+\frac{p}{2})}dz p^{2\alpha} \frac{n_B(p(z+v))+n_F(p(z+v))}{z^{1-\alpha}(z+2v)^{1-\alpha}} \nonumber \\
& \tabii + \int\limits_{\frac{1}{m}(q_f-mv)}^{\frac{1}{m}(q_f-mv-\frac{p}{2})}dz  p^{2\alpha}\frac{n_B(p(z+v))+n_F(p(z+v))}{z^{1-\alpha}(z+2v)^{1-\alpha}}   \bigg).
\end{align}
The second term is much smaller than the first term at low temperatures. We will justify that this is the case below. Dropping the second term, one finds
\beq
\mathrm{Im} \ \Sigma(T) &=& - \frac{\sin(\pi\alpha)}{2\pi^2}\frac{mu^2}{q_f}\int\limits_0^{\Lambda}dp  \int\limits_{0}^{\frac{1}{m}(q_f-mv)}dz  p^{2\alpha}\frac{n_B(p(z+v))+n_F(p(z+v))}{(z+2v)^{1-\alpha} z^{1-\alpha}}. 
\eeq
Making a change of variables $p = Tx/v$, one obtains
\beq
\mathrm{Im} \ \Sigma(T) &=& - \frac{\sin(\pi\alpha)}{2\pi^2}\frac{mu^2}{q_f v^3} T^{1+2\alpha}\int\limits_0^{v\Lambda/T}dx  \int\limits_{0}^{\frac{1}{m}(q_f-mv)}dz  \frac{x^{2\alpha}}{(\frac{z}{v}+2)^{1-\alpha} (\frac{z}{v})^{1-\alpha}}\bigg(\frac{1}{(e^{x(\frac{z}{v}+1)}-1)} + \frac{1}{(e^{x(\frac{z}{v}+1)}+1)} \bigg). 
\eeq
In the limit $T\ll v\Lambda$, if the integral over $x$ converges when the upper limit is replaced by $\infty$, we have
\beq
\mathrm{Im} \ \Sigma(T) &=&  -C_1 u^2T^{1+2\alpha}
\eeq
where $C_1$ is a constant. The error of dropping the second term in Eq. \ref{eq:sigmab1} is bounded by
\beq
|\Delta \mathrm{Im} \Sigma(T)|  < \frac{\sin(\pi\alpha)}{8\pi^2}\frac{u^2}{q_f}\int\limits_0^{\Lambda}dpp^{1+2\alpha} \bigg( \frac{n_B(\frac{pq_f}{m})+n_F(\frac{pq_f}{m})}{(\frac{q_f}{m}+v)^{1-\alpha}(\frac{q_f}{m}-v)^{1-\alpha}} + \frac{n_B(p(\frac{q_f}{m}-\frac{\Lambda}{2m}))+n_F(p(\frac{q_f}{m}-\frac{\Lambda}{2m}))}{(\frac{q_f}{m}+v-\frac{\Lambda}{2m})^{1-\alpha}(\frac{q_f}{m}-v-\frac{\Lambda}{2m})^{1-\alpha}} \bigg)
\eeq
Making a change of variables $p = \frac{mT x}{q_f}$ in the first term and $p = \frac{mT x}{(q_f-\frac{\Lambda}{2})}$ in the second term, we get
\begin{align}
|\Delta \mathrm{Im} \Sigma_B(T)| < u^2T^{2+2\alpha}\frac{\sin(\pi\alpha)}{8\pi^2 q_f}\Bigg(& \frac{(\frac{m}{q_f})^{2+2\alpha}}{(\frac{q_f}{m}+v)^{1-\alpha}(\frac{q_f}{m}-v)^{1-\alpha}}\int\limits_0^{\frac{q_f\Lambda}{mT}}dx x^{1+2\alpha}(\frac{1}{e^{x}-1}+\frac{1}{e^{x}+1}) \nonumber \\
&+ \frac{(\frac{m}{q_f-\frac{\Lambda}{2m}})^{2+2\alpha}}{(\frac{q_f}{m}+v-\frac{\Lambda}{2m})^{1-\alpha}(\frac{q_f}{m}-v-\frac{\Lambda}{2m})^{1-\alpha}}\int\limits_0^{\frac{(q_f-\frac{\Lambda}{2})\Lambda}{mT}}dx x^{1+2\alpha}(\frac{1}{e^{x}-1}+\frac{1}{e^{x}+1}) \Bigg) 
\end{align}
In the small $T$ limit, i.e. $T \ll \frac{q_f\Lambda}{m}$ and $T \ll (\frac{q_f}{m}-\frac{\Lambda}{2m})\Lambda$, we find that
\beq
|\Delta\mathrm{Im}\Sigma(T)| < C_2 u^2T^{2+2\alpha}
\eeq
where $C_2$ is a constant. Hence,  $|\mathrm{Im}\Sigma| \gg |\Delta\mathrm{Im}\Sigma|$. This result justifies the omission of the second term in Eq. \ref{eq:sigmab1} at low temperatures.

The analogous expression to Eq. \ref{eq:sigma_combine_final_3d} in the $d=2$ case is
\begin{align} \label{eq:sigma_combine_final_2d}
\mathrm{Im} \Sigma(\omega=0,\vec q_f) = -\frac{\sin(\pi\alpha)}{2\pi^2}\frac{mu^2}{q_f}\int\limits_{0}^{\Lambda}dp \bigg(&\int\limits_{0}^{-\frac{p}{2m}+\frac{q_f}{m}-v}dz \frac{p^{2\alpha-1}}{\sqrt{1-\frac{m^2}{q_f^2}(z+v+\frac{p}{2m})^2}}\frac{n_B(p(z+v))+n_F(p(z+v))}{z^{1-\alpha}(z+2v)^{1-\alpha}}  \nonumber \\
& + \int\limits_{0}^{\frac{p}{2m}+\frac{q_f}{m}-v}dz \frac{p^{2\alpha-1}}{\sqrt{1-\frac{m^2}{q_f^2}(z+v-\frac{p}{2m})^2}}\frac{n_B(p(z+v))+n_F(p(z+v))}{z^{1-\alpha}(z+2v)^{1-\alpha}}  \bigg).
\end{align}
Because of the factors $\frac{1}{\sqrt{1-\frac{m^2}{q_f^2}(z+v\pm\frac{p}{2m})^2}}$ in the integrands, the argument we used in the $d=3$ case cannot be applied to show that $\mathrm{Im}\Sigma \propto -T^{2\alpha}$ at low temperatures. 

%
\end{document}